\documentclass{article}
\usepackage{amsmath,amsthm,latexsym,amssymb,amsfonts,epsfig,psfrag}

\makeatletter
\@addtoreset{equation}{section}
\makeatother

\newtheorem{Conjecture}{Conjecture}[section]
\def\be{\begin{equation}}
\def\ee{\end{equation}}
\def\ba{\begin{eqnarray}}
\def\ea{\end{eqnarray}}
\def\Nl{{\mathchoice
{\setbox0=\hbox{$\displaystyle\rm N$}\hbox{\hbox to0pt
{\kern0.4\wd0\vrule height0.9\ht0\hss}\box0}}
{\setbox0=\hbox{$\textstyle\rm N$}\hbox{\hbox to0pt
{\kern0.4\wd0\vrule height0.9\ht0\hss}\box0}}
{\setbox0=\hbox{$\scriptstyle\rm N$}\hbox{\hbox to0pt
{\kern0.4\wd0\vrule height0.9\ht0\hss}\box0}}
{\setbox0=\hbox{$\scriptscriptstyle\rm N$}\hbox{\hbox to0pt
{\kern0.4\wd0\vrule height0.9\ht0\hss}\box0}}}}
\def\Zl{{\mathchoice
{\setbox0=\hbox{$\displaystyle\rm Z$}\hbox{\hbox to0pt
{\kern0.4\wd0\vrule height0.9\ht0\hss}\box0}}
{\setbox0=\hbox{$\textstyle\rm Z$}\hbox{\hbox to0pt
{\kern0.4\wd0\vrule height0.9\ht0\hss}\box0}}
{\setbox0=\hbox{$\scriptstyle\rm Z$}\hbox{\hbox to0pt
{\kern0.4\wd0\vrule height0.9\ht0\hss}\box0}}
{\setbox0=\hbox{$\scriptscriptstyle\rm Z$}\hbox{\hbox to0pt
{\kern0.4\wd0\vrule height0.9\ht0\hss}\box0}}}}
\def\Ql{{\mathchoice
{\setbox0=\hbox{$\displaystyle\rm Q$}\hbox{\hbox to0pt
{\kern0.4\wd0\vrule height0.9\ht0\hss}\box0}}
{\setbox0=\hbox{$\textstyle\rm Q$}\hbox{\hbox to0pt
{\kern0.4\wd0\vrule height0.9\ht0\hss}\box0}}
{\setbox0=\hbox{$\scriptstyle\rm Q$}\hbox{\hbox to0pt
{\kern0.4\wd0\vrule height0.9\ht0\hss}\box0}}
{\setbox0=\hbox{$\scriptscriptstyle\rm Q$}\hbox{\hbox to0pt
{\kern0.4\wd0\vrule height0.9\ht0\hss}\box0}}}}
\def\Rl{{\mathchoice
{\setbox0=\hbox{$\displaystyle\rm R$}\hbox{\hbox to0pt
{\kern0.4\wd0\vrule height0.9\ht0\hss}\box0}}
{\setbox0=\hbox{$\textstyle\rm R$}\hbox{\hbox to0pt
{\kern0.4\wd0\vrule height0.9\ht0\hss}\box0}}
{\setbox0=\hbox{$\scriptstyle\rm R$}\hbox{\hbox to0pt
{\kern0.4\wd0\vrule height0.9\ht0\hss}\box0}}
{\setbox0=\hbox{$\scriptscriptstyle\rm R$}\hbox{\hbox to0pt
{\kern0.4\wd0\vrule height0.9\ht0\hss}\box0}}}}
\def\Cl{{\mathchoice
{\setbox0=\hbox{$\displaystyle\rm C$}\hbox{\hbox to0pt
{\kern0.4\wd0\vrule height0.9\ht0\hss}\box0}}
{\setbox0=\hbox{$\textstyle\rm C$}\hbox{\hbox to0pt
{\kern0.4\wd0\vrule height0.9\ht0\hss}\box0}}
{\setbox0=\hbox{$\scriptstyle\rm C$}\hbox{\hbox to0pt
{\kern0.4\wd0\vrule height0.9\ht0\hss}\box0}}
{\setbox0=\hbox{$\scriptscriptstyle\rm C$}\hbox{\hbox to0pt
{\kern0.4\wd0\vrule height0.9\ht0\hss}\box0}}}}
\def\Hl{{\mathchoice
{\setbox0=\hbox{$\displaystyle\rm H$}\hbox{\hbox to0pt
{\kern0.4\wd0\vrule height0.9\ht0\hss}\box0}}
{\setbox0=\hbox{$\textstyle\rm H$}\hbox{\hbox to0pt
{\kern0.4\wd0\vrule height0.9\ht0\hss}\box0}}
{\setbox0=\hbox{$\scriptstyle\rm H$}\hbox{\hbox to0pt
{\kern0.4\wd0\vrule height0.9\ht0\hss}\box0}}
{\setbox0=\hbox{$\scriptscriptstyle\rm H$}\hbox{\hbox to0pt
{\kern0.4\wd0\vrule height0.9\ht0\hss}\box0}}}}
\def\Ol{{\mathchoice
{\setbox0=\hbox{$\displaystyle\rm O$}\hbox{\hbox to0pt
{\kern0.4\wd0\vrule height0.9\ht0\hss}\box0}}
{\setbox0=\hbox{$\textstyle\rm O$}\hbox{\hbox to0pt
{\kern0.4\wd0\vrule height0.9\ht0\hss}\box0}}
{\setbox0=\hbox{$\scriptstyle\rm O$}\hbox{\hbox to0pt
{\kern0.4\wd0\vrule height0.9\ht0\hss}\box0}}
{\setbox0=\hbox{$\scriptscriptstyle\rm O$}\hbox{\hbox to0pt
{\kern0.4\wd0\vrule height0.9\ht0\hss}\box0}}}}

\title{\bf The Holst Spin Foam Model via Cubulations}

\author{
Aristide Baratin$^1$\thanks{{aristide.baratin@aei.mpg.de}},
Cecilia Flori$^2$\thanks{{cflori@perimeterinstitute.ca}},
Thomas Thiemann$^{3}$\thanks{{thiemann@aei.mpg.de}}
\\
\\
{\small  ${}^1$ Max Planck Institute for Gravitational Physics, Albert-Einstein-Institute,} \\
{\small Am M\"uhlenberg 1, 14476 Potsdam (Germany)} \\
{\small ${}^2$ Perimeter Institute for Theoretical Physics,}\\ 
{\small 31 Caroline Street N, ON N2L 2Y5, Waterloo (Canada)}\\
{\small ${}^3$ Institute for Theoretical Physics III, University of Erlangen-N\"urnberg} \\
{\small Staudtstrasse 7, 91058 Erlangen (Germany)}
}

\date{{\small Preprint AEI-2008-093}}

\begin{document}

\maketitle

\begin{abstract}

Spin foam models are an  attempt for a covariant, or path integral formulation of canonical loop quantum gravity. 
The construction of such models usually rely on the Plebanski formulation of general relativity as a constrained BF theory and is    
based on the discretization of the action on a simplicial triangulation,  which may be viewed as an ultraviolet regulator.  
The triangulation dependence can be removed by means of group field theory techniques, which allows one to sum over all triangulations.  
%Subtle tasks for these models  lie on 
The main tasks for these models are the correct quantum 
implementation of the Plebanski constraints, the existence of a semiclassical
sector implementing additional `Regge-like' constraints  arising from simplicial  
triangulations, and the definition of the physical inner product of loop quantum gravity via group field theory.
Here we propose a new approach to tackle these issues stemming directly from the {\sl Holst action} for general relativity, which
is also a proper starting point for canonical loop quantum gravity. The discretization is performed by means of a `cubulation' of the manifold rather than a triangulation. We give  a direct interpretation of the resulting spin foam model as a generating functional for the 
$n$-point functions on the physical Hilbert space at finite regulator.  
This paper focuses on ideas and tasks to be performed before the model can 
be taken seriously. However, our analysis reveals some interesting features of this model: first, the structure of its amplitudes 
differs from the standard spin foam models. Second, the tetrad $n$-point functions admit a `Wick-like' structure. 
Third, the restriction to simple representations does not automatically occur -- unless one makes use of the time gauge,  just as in the classical theory.

\end{abstract}

\section{Introduction}
\label{s1}

Spin foam models (SFM)\cite{1} are an attempt at a covariant or path 
integral 
formulation of canonical Loop Quantum Gravity (LQG) \cite{2,3,3a}. In 
their
current formulation (for e.g \cite{13, 15, 16, DanieleAristide, EteraMaite}), SFM exploit the Plebanski formulation \cite{4} of 
pure General Relativity (GR) as a constrained BF theory. This approach 
is 
well motivated because one can view the Plebanski action as a 
kind of perturbation of the BF action (albeit the perturbation parameter
is a Lagrange multiplier field which one needs to integrate over in a 
path integral). The path integral for BF theory, however, is under 
good control \cite{5} so that one may hope to get a valid path integral
formulation for GR by functional current derivation methods \cite{6}
familiar from ordinary QFT. 

As we will try to explain in the next section (see also \cite{3}) the 
quantum implementation of the 
so called simplicity constraints of Plebanski theory, to the best 
knowledge of the authors, has still not been achieved to full 
satisfaction from first principles in these models. 
They are called simplicity constraints 
because they enforce the B field of BF theory to be simple, that is, to 
originate from a tetrad. Clearly, unless the simplicity constraints 
are properly implemented, the resulting theory has little to do with 
quantum gravity. An issue to keep in mind is that the solutions to 
the classical simplicity constraint consist of five sectors, two of 
which give rise to $\pm$ times the Palatini action, two of which give 
rise to $\pm$ times a topological action and a degenerate sector. All
of these sectors are a priori included in a sum over Plebanski histories which 
may or may not be what one wants\footnote{For a recent proposal to tackle this issue, see  \cite{Engle}.}.

It is appropriate to mention also further constraints in SFM at this 
point. The construction of these models relies on a simplicial 
triangulation $\tau$ of the differential 4-manifold as well as a dual 
graph $\tau^\ast$. A recent analysis has shown \cite{6a} that freely specifying geometrical 
data (areas or fluxes)
on the faces of $\tau$  tends to lead to inconsistent values of 
the lengths of the edges of $\tau$ unless so-called Regge constraints are imposed,  in 
addition to the  simplicity constraints. These constraints are important to 
be taken care of if one wants to relate SFM to the 
established theory of Regge calculus \cite{6aa} and in order to capture the correct 
semiclassical limit\footnote{Note however that there is no reason  to require such additional constraints 
to be implemented in the strict context of canonical loop quantum gravity, where the holonomy, flux, area, triad or length operators labelled by 
curves or surfaces have no direct physical meaning (only do their occurence in compound operators 
assembeled from them and which are Dirac observables or 
constraint operators. Moreover  in general these curves and surfaces do not even relate to any simplicial structure, so there is no  
triangulation with respect to which one would be interested in 
relating the lengths of the edges of its 1 - skeleton to the areas of 
its surfaces (in fact  in order to establish the relation between LQG and SFM it 
would seem that one needs to include spin network states on all possible
boundary graphs into the SFM analysis  -- except if one follows 
the philosophy of \cite{AQG}). Therefore Regge-like constraints never occur  in LQG.}:
in fact Regge calculus is formulated directly in terms 
of edge lengths while in SFM one rather works with electrical fluxes 
or areas; but a typical simplicial triangulation has far more faces than edges in 
$\tau$ so that assigning a length to an edge from given area values 
may be ambiguous and/or inconsistent. Note that in our approach, on the other hand, since the path 
integral is explicitly based on the Holst action, there is no necessity 
to relate it to the Regge action -- which for Plebanski's theory is of 
course a challenge.

%The authors of \cite{6a} argue that this may pose a problem 
%also for canonical LQG and that the Hamiltonian constraint of canonical
%LQG should maybe commute with these Regge constraints.
%In order avoid further confusion in the literature on this puzzling
%issue, we will take here the opportunity to show that
%{\it there are no Regge like constraints in canonical LQG to be taken 
%care of.} 

In fact, one possibility to make progress on the common issues of the standard formulation of spin foam models,  is based on a  
very simple idea which, to the best knowledge of the authors, occurs for 
the first time in \cite{7a}: Namely, simply try to formulate the path 
integral in terms of the Holst action \cite{7} rather than the Plebanski 
action. Not only is the Holst action a valid starting point for 
canonical LQG, but also the simplicity constraints are explicitly solved 
in that one works entirely with tetrads from the beginning. 
More 
precisely, the Holst 
action uses a specific quadratic expression in the tetrads for the B 
field of BF theory which also depends on the Immirzi parameter 
\cite{8}. Hence, the Holst action depends on a specific, non degenerate 
linear 
combination of the four non degenerate solutions of the simplicity 
constraints (see next section for details) and is thus at the same time 
more general and more restricted because the Holst path integral will 
not sum over the afore mentioned five sectors of Plebanski's theory. 
As already mentioned, it is at present 
debated how the fact that one actually takes a sum over all histories 
with a mixture of positive and negative Palatini and topological actions 
affects the semiclassical properties of the Plebanski path integral.

As observed in \cite{7a}, since the Holst action is quadratic in the 
tetrads, one can in principle integrate out the tetrad in the resulting 
Gaussian integral. This has been 
sketched in \cite{7a}, however, the expressions given there are far from 
rigorous. Here we will give a rigorous expression. Also, we will include 
the correct measure factor \cite{8a} resulting from the second class 
constraints 
involved in the Holst action and making sure that the path integral 
qualifies as a reduced phase space quantisation of the theory,  as it has 
been stressed in \cite{8b}. A similar analysis has been carried out 
for the Plebanski theory in \cite{8c}, however, the resulting measure 
factor is widely ignored in the SFM literature\footnote{For a recent review on the relation between spin foam models and canonical quantization, see \cite{ReviewKarim}.}. 
The result of the 
Gaussian integral is an interesting determinant that displays the full
non linearity of Einstein's theory. When translating the remaining 
integral over the connection in the partition function into SFM 
language, that is, sums over 
vertex, edge and face representations, one sees that our model differs 
drastically from all current models. 

Of course, we also need to introduce an IR and UV regulator in the form 
of a finite cell decomposition. Two observations lead us to depart 
from the usual SFM approach where one works with simplicial cell 
complexes.
The first one is the result \cite{6h} which demonstrates that 
current semiclassical states used in LQG do not assign good classical 
behaviour to the volume operator \cite{8d} of LQG unless the underlying 
graph has 
cubic topology (see also \cite{6g,AQG}).   Since the volume operator plays a pivotal role for 
LQG as it defines triad operators and hence the dynamics, this is a 
first motivation to consider cubic  triangulations of the four manifold, which we coin ``cubulations'' (see for e.g \cite{9} and references therein).  
Note that the result of \cite{6h} implies that current spin foam models based on simplicial cell complexes
do not admit the semiclassical states \cite{6f} as boundary 
states which could mean that the current models maybe have to be 
extended to  more general triangulations.
The second observation is that the original motivation for considering 
simplicial 
cell complexes in current SFM comes from their closeness to BF theory.
BF theory is a topological quantum field theory (TQFT)  and therefore one would like to keep 
triangulation independence of the BF SFM amplitude. That this is 
actually true is a celebrated result in BF theory. In 
particular, in order to keep triangulation independence it is 
necessary to integrate the B field over the triangles $t$ of the 
tringulation and the F field over the faces $f$ bounding the loops in a 
dual graph
\cite{9}. However, GR is not a TQFT and therefore the requirement to 
have triangulation independence is somewhat unclear. Of course it is 
natural if one wants to exploit the properties of BF theory but not if 
one takes a different route as we tend to do here. Hence, if we drop
that requirement, then it is much more natural to refrain from 
considering the dual graph in addition to the triangulation. 
Working with cubulations also greatly simplifies the realization of gauge invariance in discrete models. 
In fact gauge invariance is related to the closure constraint in SFM which is a 
subtle issue, as we will see in the next section. If one 
works just with a triangulation and drops the dual graph then such  issues 
are easy to take care of.  Finally, the use of cubulations also fits nicely   with the framework of Algebraic 
Quantum Gravity \cite{AQG} which in its minimal version also is 
formulated in terms of algebraic graphs of cubic topology only. \\

%This also 
%solves another issue:
%Notice that the gauge group $SO(p,q)$ acts on the B field of BF theory 
%by the  adjoint  action and on the connection $A$ underlying $F$ in the usual way. The 
%question is where the gauge transformation acts on the discretised 
%variables $B(t),\;A(\partial f)$ (flux and holonomy). It would be 
%natural to have the gauge group act at the barycentres of $t$ and at 
%the starting point of the loop $\partial f$ which will be a vertex of 
%the dual graph. However, the vertices of the dual graph and 
%the triangles are disjoint from each other, the edges of the graph are 
%dual to the tetrahedra of the cell complex. Hence, 
%{\it local} gauge invariance in discretised BF theory at the level of 
%the action is not manifest and even less in Plebanski theory. In fact gauge 
%invariance is related to the closure constraint in SFM which is a 
%subtle issue as we will see in the next section. If one 
%works just with 
%a triangulation and drops the dual graph then gauge invariance issues 
%are 
%easy to take care of. 
%Hence, it is motivated to work with a triangulation that maximally 
%simplifies the Gaussian integral. As we will show, this again leads to 
%cubulations. This also nicely fits with the framework of Algebraic 
%Quantum Gravity \cite{AQG} which in its minimal version also is 
%formulated in terms of algebraic graphs of cubic topology only.\\
%\\
The architecture of this article is as follows:
\\

In section two we give a non technical review of current spin foam 
models. We sketch their derivation from the classical Plebanski action 
focusing on the points where a first principle argument is missing.
These issues will be the motivation for our different route.

%In section three we clarify the issue mentioned above, namely whether
%the Regge constraints on SFM discovered recently in \cite{6a} also 
%arise in the canonical formulation. We prove that they do not play
%any role in LQG.

In section 
three 
we derive the Holst spin foam model using cubulations as UV 
regulator as motivated above. As this is a exploratory paper only,
we will not worry about convergence issues which will be properly 
addressed in subsequent works. More precisely, what we compute are 
tetrad n-point functions. These should contain sufficient information 
to compute anything of interest in LQG such as graviton scattering 
amplitudes as in \cite{9a} via LSZ (Lehmann -- Zimmermann -- Symanzik) 
like formulas as in ordinary QFT \cite{Haag} which allows to 
reconstruct the S -- matrix from symmetric vacuum n -- point functions. 
Of course, how these n-point functions 
are related to true observables in a diffeomorphism invariant theory
is a subtle issue which will be clarified in a separate paper \cite{9b}.
Here we only give a summary.
The n-point functions can be computed in closed form up to a remaining
functional integral over the connections. This can be done 
for either signature of the spacetime metric. At this point one could 
invoke SFM techniques and expand the integral using harmonic analysis 
on the gauge group. The resulting intertwiner displays a much more 
complictated structure than in any of the current spin foam models.
In particular, pictorially speaking one basic building block
is an {\it octagon diagramme} an analytic expression for which could 
be called the {\it 96 -- j symbol} in the case of $G=SO(4)$.
Yet, the n-point functions display a certain Wick like structure as 
if they came from a Gaussian integral. What makes the theory 
interacting and obstructs the tetrads from being a generalised free 
field\footnote{Roughly, a generalised free scalar field is such that all
its n-point functions are already determined by its two point function.} 
is the additional functional integral over 
the connection. In background dependent QFT the moments of a Gaussian 
measure depend on a background dependent covariance (usually depending 
on the Laplacian (in the Euclidean setting) and the mass). Our 
theory 
behaves similar, just that due to background independence the covariance 
is itsef a field that must be integrated over. This is similar in spirit 
to what happens in 3D \cite{9c} when coupling GR to point 
particles: There, when integrating over the gravitational degrees of 
freedom 
one ends up with particles moving on a non commutative geometry. Here 
instead of a non commutative geometry we obtain an interacting theory
of tetrad fields.

In section 
four
we conclude and outline the missing tasks that need to be 
performed before our model can be taken seriously. An interesting result 
of our analysis is that in the present formulation which lacks the 
simplicity constraints of Plebansk's theory, the irreducible 
representations of $Spin(p,4-p)$ are not forced to be simple. Simple 
representations, which basically reduce $Spin(p,4-p)$ to an $SU(2)$ 
subgroup, can only arise if we impose the time gauge which in the 
classical theory is used in order to reduce the Holst connection to the 
Ashtekar -- Barbero -- Immirz connection which is a necessary ingredient 
in the canonical quantisation programme. Gauge fixing conditions of 
course naturally arise in any attempt to make formal path integral 
expressions better behaved and here the situation is similar.  

Two appendices treat some simple technical aspects of this work.
 
Most parts of this paper do not depend on whether
the spacetime signature is Lorentzian or Euclidean.

\section{Outline of Current Spin Foam Models}
\label{s2}

In this section we intend to give a brief summary of the developments in
spin foam models with a focus on the derivation of the current models from the 
Plebanski action and the gaps in that derivation. This serves as the 
motivation for the present paper.\\ 
\\
To begin with, it is worth mentioning that the 
classical solutions to 
the simplicity constraints actually comprise altogether five sectors,
namely two topological sectors $B=\pm e\wedge e$, 
two Palatini sectors $B=\ast e\wedge e$ (where $\ast$ 
denotes the Hodge map with respect to the internal Minkowski or 
Euclidean metric) and one degenerate sector. In a path integral a sum
over all these sectors will occur while one would expect that one 
should only include one of the Palatini sectors or maybe a Holst
sector $B=\ast e\wedge e+\frac{1}{\gamma} e\wedge e$ \cite{7}
in order to have a path integral for Einstein's theory. Here $\gamma$ is 
the Immirzi parameter of LQG \cite{8}. 

We now sketch the usual ``derivation'' of spin foam models from the Plebanski 
action:\\ 
The Plebanski action is of the form 
\be \label{2.1}
S=\int\; Tr(B\wedge F(A)+\Phi \cdot B\wedge B)
\ee
where $\Phi$ is a scalar 
Lagrange multiplier field with values in the tensor product of two 
copies of $so(1,3)$ or $so(4)$ depending on the signature and $F$ is the 
curvature of the connection $A$. In a formal path integral formulation 
one integrates $\exp(iS)$ over $A, B, \Phi$. Integrating first over 
$\Phi$ we are left with a partition function of the form
\be \label{2.2} 
Z=\int\;[dA]\;[dB]\;\delta(C(B))\;\exp(i\int\;Tr(B\wedge F))
\ee
where $C(B)$ denotes the collection of the simplicity constraints on 
$B$. If one would solve the delta distribution by integration over $B$
one would get the afore mentioned sum over the five sectors and integral
over the tetrad fields. However, this would result in a complicated 
expression which does not exploit the relation of Plebanski's 
formulation to BF theory. Thus, rather than doing that, one notices that 
roughly speaking 
\be \label{2.3}
B \exp(i\int\;Tr(B\wedge F))=\frac{1}{i}\; \frac{\delta}{\delta F}
\exp(i\int\;Tr(B\wedge F))
\ee
Denoting the functional derivative by $X$ one can now formally pull
the $\delta$ distribution out of the $B$ integral and perform the 
integration over $B$ resulting in 
\be \label{2.4}
Z=\int\;[dA]\;\;\delta(C(X))\;\cdot\;\delta(F)
\ee
Without the ``operator'' $\delta(C(X))$ this would be the formal 
partition 
function of BF theory. Thus, one has achieved the goal to preserve 
the closeness of the theory to BF theory. One now should expand 
$\delta(F)$ in terms of eigenfunctions of the collection of operators 
$C(X)$ and keep only the zero eigenfunctions multiplied by $\delta(0)$.

In order to give meaning to those formal expressions one has 
to introduce a UV and IR cutoff as is customary in constructive QFT. 
That is, one considers finite simplicial triangulations $\tau$ of the 
(possibly compact) differential 4 manifold
and dual graphs $\tau^\ast$. The two forms $B$ are now 
approximated 
by integrals $B(t)$ of $B$ over 
triangles $t$ of $\tau$ while the curvatures $F$ are approximated by 
holonomies $A(\partial f)$ around the loops $\partial f$ of the faces 
$f$ dual to the 
triangles $t$. One writes $f(t)$ for the face dual to $t$. The BF action
is then discretised by
\be \label{2.5}
\sum_{t\in \tau}\;Tr(A(\partial f(t))\;B(t))
\ee
The reason to work with both $\tau$ and $\tau^\ast$ is that in fact
\be \label{2.6}
\int_M\; Tr(B\wedge F)=\sum_{t\in \tau}\; Tr(F(f(t)) B(t))
\ee
is {\it an exact identity} \cite{9} where $F(f)$ denotes the integral of 
$F$ 
over $f$. This is very convenient in particular for pure BF theory. The 
only approximation thus consists in replacing $F(f)$ 
by $A(\partial f)-1_G$.  

Likewise, the functional derivatives $X$ must be 
approximated by ordinary derivatives $X_t$ with respect to the 
variables $F(t):=A(\partial f(t))-A(\partial f(t))^T$. 
Notice that when defined like that, the $X_t$ are mutually commuting\footnote{
We refer to \cite{10} for the exploration of the model with this definition of $X_t$.}. 
However, this is not what is done in current models. Rather
one replaces $X_t$ by $Y_t$, the right invariant vector field on the 
copy of $G$ associated
with the variable $A(\partial f(t))$. 
Upon  spin foam quantization the discrete B variables thus become explicitly non-commutative\footnote{In fact, as shown in the recent work \cite{DanieleAristide, aristidedaniele2, Lqgflux}, 
spin foam models defined as constrained BF models take the form of non-commutative discrete path integrals making use of a star product on functionals of the B variables. 
It can also be shown that the generating group field theories are just a particular class of  non-commutative field theories \cite{GFTnoncomm}.}.
The reason for 
doing this replacement is that the $Y_t$ have a simpler action on the delta 
distribution 
\be \label{2.7}
\delta(F):=\prod_{t\in \tau} \delta_G(A(\partial f(t))
\ee
It is usually justified by saying that $\delta(F)$ has support on
$A(\partial f)=1_G$ and that $Y_t,\;X_t$ differ by multiplication with 
holonomies which should be supported at $1_G$. However, this argument is 
certainly not rigorous because 
the support of the $\delta$ distribution can drastically change when 
acting with 
differential operators. Moreover, as already said, this substitution 
comes with a price:
While the simplicity constraints in terms of $X_t$ are mutually 
commuting, those in terms of $Y_t$ are not. 
In addition, one does not impose all the 
simplicity constraints but only a subset of them:
There are three types: Constraints involving 1. the same triangle, 2. 
two
triangles sharing an edge and 3. two triangles sharing a vertex. The 
latter constraint is implied by the so called closure constraint on 
tetrahedra $T$ 
\be \label{2.8}
\sum_{t\in T} \; Y_t=0
\ee
(but not vice versa).
This constraint looks as if it would be automatically satisfied because 
it looks like a gauge invariance condition. However, the product of 
$\delta$ distributions (\ref{2.7}) in $\Delta(F)$ is not annihilated by 
the closure 
constraint (\ref{2.8})! This is obvious from the fact that the product 
of $\delta$ 
distributions involves products of the form 
\be \label{2.9}
\prod_{t\in T} \chi_{\pi_t}(A(\partial f(t))
\ee
where $\pi$ denotes an irreducuible representation of $G$ and $\chi_\pi$
its character. However, there is no gauge invariant interwiner among the 
loops $\partial f(t),\;t\in T$. One usually argues that the closure 
constraint is taken into account because {\it after} integrating over 
$A$ one is only left with gauge invariant intertwiners, but strictly this is 
wrong {\it  
before} integrating\footnote{See however \cite{FreidelConrady, aristidedaniele2, DanieleAristide} for discussions of this point.}. 
In fact since integration with the Haar measure 
always projects out the gauge invariant part, anything can be made gauge 
invariant this way. We feel that  
neglecting the 3rd kind of simplicity constraint (implied by taking 
the closure constraint for granted) makes the model too local. 
The effect of truly taking the closure constraint into 
account is also  explored in \cite{10}.

As already said, even the simplicity constraints of the first two 
types are anomalous as they imply vanishing volume 
\cite{4,3} and fix the above mentioned intertwiner to be unique 
(the model has not enough degrees of freedom). This 
and other investigations has ruled out\footnote{See however \cite{aristidedaniele2} for a recent critical review of the various arguments raised against the Barrett-Crane model} the Barrett -- 
Crane model \cite{5} which however was an important step in the 
research in spin foam models because it triggered the development of model 
independent mathematical tools. 
Recent activities in spin foam models therefore focussed
on trying to implement the simplicity constraints of the first two types
differently. Thus, for instance, the work \cite{13} one uses Master 
Constraint type of techniques which were developed in a different 
context \cite{14} in order to treat second class
constraints via a Gupta-Bleuer quantization procedure. In \cite{15}, one refrains from imposing the  
simplicity constraints as operator conditions altogether but rather 
imposes them semiclassically by expanding spin foam amplitudes in terms of 
group coherent states \cite{16} developed by Perelomov \cite{17} and 
then uses the simplicity condition on the classical bivectors on which 
the semiclassical amplitudes depend. More recently,  the work \cite{EteraMaite} exploits a spinorial representation of spin network states to implement a Gupta-Bleuer quantization of the simplicity constraints, shown to be  solved exactly by coherent  states with appropriate labels. 
Finally, \cite{GFTnoncomm, DanieleAristide} exploit a non-commutative metric representation of spin network states and a non-commutative simplicial path integral representation of quantum BF theory  to implement the simplicity constraints as strong constraints on the discrete (and non-commutative) B variables.

Some of these methods give rise to models with better semiclassical properties \cite{18} and to better ways to disentangle  the topological from the Palatini sector. However in our opinion  a satisfactory derivation from first principles is still missing. By this we mean that one should be able to arrive at those 
models starting from the Plebanski action, another classically equivalent action or the Hamiltonian formulation and then carry out
integrations and imposition of constraints without intermediate 
approximations or ad hoc substitutions as those listed above\footnote{There has been recent  work \cite{Bonzom, 10, DanieleAristide} in this direction, where simplicity constraints are clearly implemented in the measure of a path integral. The  novelty of the present approach, however,  is to start directly from the Holst gravity action, which avoids to have to deal with simplicity constraints to begin with.}. 
This is precisely the motivation of the current paper.

\section{Derivation of the Model}
\label{s5}

This is the main section of the paper. It is subdivided into
five parts. In the first we motivate the use of {\sl cubulations} from 
different perspectives and discuss some of their properties. In the 
second we sketch the relation between path integral n-point functions 
and physical (observable) correlators in terms of the physical inner 
product of the theory. More details on that issue are given in
\cite{9b}. This crucially works via a choice of gauge fixing or clock 
system. In the third part we apply our machinery to n-point tetrad 
functions or equivalently to a 
generating function of a (complex, regulated) measure. This measure 
displays a Gaussian-like structure and we can accordingly integrate out 
half of the degrees of freedom under some assumptions about the choice 
of gauge fixing. In the fourth part we discuss the 
properties of the resulting integral over the remaining degrees of 
freedom, its Wick-like structure and the structure of the vertex 
amplitude of the corresponding spin foam model obtained upon harmonic analysis 
on the gauge group. Finally, in the fifth part we discuss how these 
n-point functions are related to the physical inner product and the 
kinematical Hilbert space of LQG, in particular, how the covariant 
connection of the Holst path integral reduces to the Ashtekar -- Barbero 
-- Immirzi connection of the canonical theory in physical amplitudes.

\subsection{Cubulations}
\label{s5.1}

In contrast to the standard way to discretize the theory using simplicial triangulations, our approach will be based on {\sl cubulations} of the underlying manifold. 
The advantages of these, spelled out below, are:
\begin{itemize}
\item To facilitate gauge invariant discretization of the classical theory 
\item To insure the existence of a semi-classical sector within the boundary Hilbert space \cite{6h}
\end{itemize}
but the main advantage is a practical one: 
\begin{itemize}
\item To  permit a discretization of the action in terms of a Gaussian with block diagonal kernel, which allows explicit computation 
of the Gaussian integrals in the partition function
\end{itemize}

\subsubsection{Gauge invariance}
\label{s5.1.1}

Let us look more closely at the issue of gauge 
invariance for BF theory which makes use also of a dual graph. Here 
gauge invariance is not 
preserved 
locally (i.e. triangle wise) in the 
formula
$\int Tr(B\wedge F)=\sum_t Tr(B(t)F(f(t))$ if both $B$ and $F$ transform 
locally in the adjoint representation. In order to make the gauge 
transformations more local, one could discretise them. 
To see how this can be achieved, recall 
that by 
definition of a cell dual to a simplex\footnote{As usual \cite{9}, an 
n-simplex is denoted by $[p_0,..,p_n]$ where the points $p_i$ denote its
corners.}  
in a 
simplicial complex $\tau$, 
the face 
$f(t)$ is a union of triangles $[\hat{t},\hat{T},\hat{\sigma}]$ 
subject to the condition $t\subset \partial T,\;T\subset \partial \sigma$.
Here 
$\hat{(.)}$ denotes the barycentre \cite{9} of a simplex and 
$T,\;\sigma$ denote 
tetrahedra and four simplicies in $\tau$ respectively. So we see that both 
$t$ and $f(t)$ contain the barycentre $\hat{t}$ in their intersection
and we could {\it define} a disjoint action of the gauge group on
both $B(t),\;F(f(t))$ at $\hat{t}$. However, this is no longer 
possible when  
using the approximation
$\sum_t Tr(B(t) A(\partial f(t)))$ because now the only natural action 
of the gauge group on the loop holonomy is by adjoint action at a 
starting point on $\partial f(t)$. Now $\partial f(t)$ is a composition of 
the half edges 
$[\hat{T},\hat{\sigma}]$ where $t\subset \partial T,\; T\subset \partial 
\sigma$ but the fundamental degrees of freedom are the holonomies along 
the edges $e=[\hat{\sigma},\hat{\sigma'}]$ for $\sigma\cap 
\sigma'=T,\;t\subset \partial T$. Obviously, the only natural starting 
point of the loops is then at the vertices $\hat{\sigma}$ which are 
disjoint from the triangles $t$. But the triangles are also disjoint 
from the half edges as a simple calculation reveals.  
To maintain gauge invariance one has to come up with a more complicated discretized action 
(for e.g in terms of wedge variables related to each other by additional holonomy variables \cite{18}). 
Such complications come from the 
fact that one is dealing simultaneously with a (simplicial) complex and 
its dual cell complex; we take this as a further piece of motivation to work only with 
the triangulation. 	

\subsubsection{Cubulations versus simplicial triangulations}
\label{s5.1.2}

The previous considerations do not specify the type of triangulations to 
be considered. As already said, the first piece of information why to 
use cubulations rather than simplicial triangulations is because the 
boundary graphs must contain cubical ones in order to make sure 
that the corresponding boundary Hilbert space contains enough 
semiclassical states \cite{6h}. However, there is an additional, 
more 
practical motivation for doing so which we discuss now.

Recall that the Holst action is given by 
\be \label{5.1}
S=-\frac{1}{\kappa}\int_M\; {\rm Tr}(G[A]\wedge e\wedge e)
=\frac{1}{\kappa}\int_M\; G_{IJ}[A]\wedge e^I\wedge e^J
\ee
Here $\kappa$ denotes Newton's constant,
\be \label{5.2}
G[A]=2(\ast F[A]+\frac{1}{\gamma} F[A])
\ee
where 
$F_{IJ}=dA_{IJ}+A_{IK}\wedge A^K\;_{J}$ denotes the curvature of 
the connection $A$, $\gamma$ is the Immirzi parameter, $\ast$ denotes 
the internal Hodge dual, that is, 
\be \label{5.3}
(\ast T)_{IJ}:=\frac{1}{2} \epsilon_{IJKL} \eta^{KM} \eta^{LN} T_{MN}
\ee
where $I,J,K,..=0,..,3$ and $\eta$ is the Minkowski or Euclidian metric
for structure group $G=SO(1,3)$ or $G=SO(4)$ respectively. As motivated 
in 
the 
introduction, we plan to keep the co -- tetrad one forms $e^I$ rather 
than introducing a B field and thus 
the simplicity constraints are manifestly solved. Moreover, the issue 
raised in \cite{6a} is circumvented as co-tetrads are labelled by curves 
and not by (overcomplete) surfaces.

In order to give 
meaning to a path integral formulation we consider a UV cutoff in terms 
of a triangulation $\tau$ of $M$ which we choose to be finite, thereby 
introducing an IR regulator as well. Let us denote the two -- 
dimensional faces of $\tau$ by $f$ and the one dimensional edges of 
$\tau$ by $l$. We want to discretise (\ref{5.1}) in a manifestly (and 
locally) gauge invariant way, just using edges and faces. To do so we 
equip all edges with an orientation once and for all. Given an edge
$l$ consider
\be \label{5.4}
e^I_l:=\int_l \; [A(l(x))]^I\;_J e^J(x)
\ee
Here $l(x)$ for $x\in l$ denotes the segment of $l$ that starts at the 
starting point of $l$ and ends at $x$ and $[A(p)]^I\;J$ denotes the 
G valued holonomy
of $A$ along a path $p$. Evidently, under 
local gauge transformations
$g:\;M\to G$, 
(\ref{5.4}) transforms as $e^I_l\mapsto g^I\;_J(b(l))\; e^J_l$ where 
$b(l)$ denotes the beginning point of $l$. 

To avoid confusion, here $g\in G$ means the following: The 
fundamental objects are the matrices $g^I\;_J$. Set 
$\tilde{g}_{IJ}:=\eta_{IK}\; g^K\;_J$. Then $g\in G$ iff  
$\tilde{g}_{IK} \tilde{g}_{JL}\eta^{KL}=\eta_{IJ}$. This is 
equivalent with $(g^{-1})^I\;_{J}=\eta^{IL} g^K\;_L \eta_{KJ}$. In 
other words
\be \label{5.5}
\widetilde{(g^{-1})}=(\tilde{g})^T
\ee
If $g^I\;_J=[\exp(F)]^I\;_J$ 
for 
some generator $F^I\;_J$ then (\ref{5.5}) means that 
$\tilde{F}_{IJ}+\tilde{F}_{JI}=0$. In abuse of notation one 
usually uses the same symbols $g,F$ and $\tilde{g},\tilde{F}$  
respectively but unless we are in the Euclidian regime we should pay 
attention to the index position.

Clearly, the curvature $F$ 
must be discretised in terms of the holonomy of $A$ along the closed 
loops $\partial f$ where we have also equipped the faces $f$ with an 
orientation once and for all. We have  
\ba \label{5.6}
F_{IJ}(f) 
&:=& \frac{1}{2}([\widetilde{A(\partial f)}]_{IJ}-
[\widetilde{(A(\partial 
f))^{-1}}]_{IJ})
\nonumber\\
&=& \frac{1}{2}([\widetilde{A(\partial f)}]_{IJ}-
[(\widetilde{(A(\partial 
f)})^T]_{IJ})
\nonumber\\
&=& \widetilde{A(\partial f)}_{[IJ]}
\nonumber\\
&\approx& \int_f \; F_{IJ}(x)
\ea
where we have used the non Abelian Stokes theorem for ``small'' 
loops, that is 
\be \label{5.7}
A(\partial f)\approx \exp(\int_f F)
\ee
and we have written $\tilde{F}_{IJ}(x):=F_{IJ}(x)$.
We may now define the antisymmetric matrix
\be \label{5.8}
G_{IJ}(f)=(\ast F(f))_{IJ}+\frac{1}{\gamma} F_{IJ}(f) 
\ee

Imagine now that we would use a simplicial triangulation. Hence $M$ is 
a disjoint (up to common tetrahedra) union of four simplicices 
$\sigma=[p_0(\sigma),..,p_4(\sigma)]$. For each $p_j(\sigma)$ label the 
four boundary edges of $\sigma$ starting at $p_j(\sigma)$ by 
$l_\mu^j(\sigma)$ and let the face (triangle) of $\sigma$ spanned by  
$l_\mu^j(\sigma),\;\;l_\nu^j(\sigma)$ be denoted by 
$f_{\mu\nu}^j(\sigma)$ with the convention 
$f_{\mu\nu}^j(\sigma)=-f_{\nu\mu}^j(\sigma)$. Now the 
orientation of $l_\mu^j(\sigma)$ either coincides with the given 
orientation of the corresponding edge in $\sigma$ or it does not. 
In the former case define $e_\mu^{I j}(\sigma):=e^I_{l_\mu^j(\sigma)}$ 
while in the latter we define  
$e_\mu^{I j}(\sigma):=[A(l_\mu^j(\sigma))^{-1} e_{l_\mu^j(\sigma)}]^I$.
Then we have 
\ba \label{5.9}
\kappa S &=& -\sum_{\sigma\in \tau}  \int_\sigma {\rm Tr}(G\wedge 
e\wedge e)
\nonumber\\
&\approx& 
\frac{1}{5}\sum_{\sigma\in \tau}\sum_{j=0}^4
\epsilon^{\mu\nu\rho\lambda} G_{IJ}(f_{\mu\nu}^j(\sigma)) 
e_\rho^{I j}(\sigma) \; e_\lambda^{J j}(\sigma)
\nonumber\\
&=:& \sum_{l,l'} G_{IJ}\;^{l,l'}\; e^I_l\; e^J_{l'}
\ea
where we have averaged over the corners of a 4 -- simplex.
For any simplicial triangulation the (symmetric in the compound 
index $(I,l)$) matrix 
$G_{IJ}^{ll'}$ is 
difficult
to present explicitly due to bookkeeping problems, even if we refrain from 
averaging over the five corners of a 4 -- simplex. Moreover, as we 
intend to perform a Gaussian integral over the $e^I_l$, we need the 
determinant of that 
matrix which is impossible to compute explicitly unless it is  
block diagonal in some sense.

The latter observation points to a possible solution. First of all
any manifold admits a cubulation, that is, a triangulation by embedded 
hypercubes\footnote{An easy proof uses the fact that every manifold can be 
triangulated by simplices. Given a D -- simplex, consider the barycentre 
of each of its ${D+1 \choose p+1}$ sub -- $p$ -- simplices for $p=0,..,D$.
Connect the barycentre of any $p+1$ -- simplex with the barycentres of 
the $p$ -- simplices in its boundary. It is not difficult to see that
this defines a cubulation of the D -- simplex and that all p -- cubes
thus defined are the same ones in common q -- simplices of the original 
simplicial complex. In other words, every simplicial complex has a 
cubulated refinement.}  \cite{8e}. 
We assume that $M$ has a countable cover 
by open sets $O_\alpha$. Consider a stratification 
by 4D regions $S_\alpha$ subordinate to it. Then $S_\alpha$ admits a 
regular cubulation, that is, the 1 -- skeleton of the cubulation of $M$ 
restricted to 
$S_\alpha$ can be chosen to be a regular cubic lattice. Non trivial 
departures from the regular cubulation only appear at the boundaries of 
the $S_\alpha$. We restrict attention to those $M$ admitting a 
cubulation such that in every compact submanifold the ratio of the 
number of cubes 
involved in the 
non -- regular regions divided by the number of cubes involved in the  
regular regions converges to zero as we take the cubulation to the 
continuum. For those $M$, up to corrections which vanish in the 
continuum limit we can treat $M$ as if it would admit a global, regular 
cubulation. 

Given a regular cubulation $\tau$, consider its set of vertices. 
In 4D, each vertex $v$ is eight valent and there are four pairs of edges 
such that the members of each pair are analytic continuations of each 
other while the tangents at $v$ of four members from mutualy different 
pairs 
are linearly independent of each other. It is therefore possible to 
assign to each edge a direction $\mu=0,1,2,3$ and an orientation such 
that adjacent 
edges in the same direction have a common analytic continuation and 
agree in their orientation. We label the edges starting at $v$ in $\mu$
direction by $l_\mu(v)$. Notice that this labelling exhausts all 
possible edges and unambiguously assigns an orientation to all of them.
The discretised co -- tetrad is then given by
\be \label{5.10}
e^I_\mu(v):=e^I_{l_\mu(v)}
\ee
Notice that the hypercubic lattice that results solves all our bookkeeping 
problems since we now may label each vertex by a point in $\mathbb{Z}^4$.

Next, given a vertex $v$ we denote by $v\pm \hat{\mu}$ the next 
neighbour vertex in $\mu$ direction. We define the plaquette loop
in the $\mu,\nu$ plane at $v$ by
\be \label{5.11}
\partial f_{\mu\nu}(v):=l_\mu(v) \circ l_\nu(v+\hat{\mu}) \circ 
l_\mu(v+\hat{\nu})^{-1} \circ l_\nu(v)^{-1}
\ee
so that $\partial f_{\nu\mu}(v)=[\partial f_{\mu\nu}(v)]^{-1}$. Notice 
that again this labelling exhausts all minimal loops in the one skeleton 
of $\tau$.
The discretised ``curvature'' is therefore 
\be \label{5.12a}
G_{IJ}^{\mu\nu}(v):=\epsilon^{\mu\nu\rho\sigma} G_{IJ}(f_{\rho\sigma}(v))
\ee

Denoting by $\sigma$ the 4D hypercubes in $\tau$ we notice that 
there is a one to one correspondence between the vertices $v$ in the 0 -- 
skeleton of $\tau$ and the hypercubes given by assigning to $\sigma$
that corner $v=(z_0,..,z_3)$ of $\sigma$ with smallest values of all
$z_0,..,z_3\in \mathbb{Z}$. We then find 
\ba \label{5.12}
\kappa S &=& \sum_{\sigma} \int_M \; G_{IJ} \wedge e^I\wedge e^J
\nonumber\\
&\approx & \sum_v \sum_{I,J,\mu,\nu} \; G_{IJ}^{\mu\nu}(v)\; 
e^I_\mu(v)\; e^J_\nu(v)
\ea
The crucial observation is now the following: Assemble pairs of indices 
into a joint index $A=(I,\mu),\;B=(J,\nu)$ etc. and let 
$e^A(v):=e^I_\mu(v),\;G_{AB}(v):=G^{\mu\nu}_{IJ}(v)$ etc..
Notice that by construction $G_{AB}(v)=G_{BA}(v)$ for all $v$. Then 
(\ref{5.12}) can be written as 
\be \label{5.13}
\kappa S \approx \sum_v \; e^T(v)\; G(v)\; e(v)
\ee
This means that using (regular) cubulations indeed the matrix
$G^{ll'}_{IJ}$ becomes block diagonal where each block is labelled by a 
vertex and corresponds to the symmetric 16 x 16 matrix $G(v)$.
This is what makes the computation of the detrminant of the huge matrix
with entries $G^{ll'}_{IJ}$ practically possible. As we will see,
the matrices $G(v)$ have a lot of intriguing symmetries which makes the 
computation of their determinant an interesting task.\\ 
\\
Interesting questions that arise in algebraic topology and which we intend 
to address in future publications are:
\begin{itemize}
\item[1.] Given any D -- cubulation, does there exist a cubulated 
refinement such that one can consistently assign to every D cube 
$\sigma$ a vertex $v$ and to all edges an orientation such that there are 
precisely D edges outgoing from $v$? We call cubulations for which 
this is possible {\it regular}. If that would be the case, we could 
generalise our discretisation from regular hypercubic lattices to 
arbitrary cubic ones 
and thus would not have to make any error at the boundaries of the 
stratified regions mentioned above. 
\item[2.] If the answer to [1.] is negative, can one choose maximally 
regular cubulations as to minimize the error in our assumption of globally 
regular cubulations? In 3D some results on that issue seem to exist 
\cite{8e}.
\item[3.] Given maximally regular cubulations, can one make an error 
estimate resulting from the neglection of the non -- trivial topology?
\end{itemize}

\subsection{Notes on n-point functions}
\label{s5.2}

In the spin foam literature the first task that one addresses is the computation 
of the partition function. However, the partition function itself has 
no obvious physical meaning even if one imposes boundary conditions on 
the paths
(spin foams) to be integrated (summed) over. The hope is that  
SFM provide a formula for the physical inner product
of the underlying constrained canonical theory which starts from some 
kinematical Hilbert space $\cal H$. The purpose of this section is to 
sketch the connection between path integrals and n -- point functions 
for a general constrained theory. We will use reduced phase space 
quantisation as our starting point. The connection with operator 
constraint quantisation and group averaging \cite{21} and more details 
can be found in \cite{9b}.\\
\\
We assume that we are given a classical theory with first class 
constraints $\{F\}$ and possibly 
second class 
constraints $\{S\}$. We turn the system into a purely second class 
system by supplementing $\{F\}$ with suitable gauge fixing conditions 
$\{G\}$. The canonical Hamiltonian $H_c$ is a linear combination of the
primary constraints plus a piece $H'_0$ non -- 
vanishing on the constraint surface of the primary constraints (it 
could be identically zero). 
It can also be written as a first class piece $H_0$ and (some of) the 
first class constraints $F$. The 
gauge fixing conditions fix the Lagrange multipliers involved in the 
canonical Hamiltonian. One may split the complete set of canonical 
pairs $(q,p)$ on the full phase space into two sets $(\phi,\pi),\;(Q,P)$
such that one can solve the system $S=F=G=0$ which 
defines the constraint surface for 
$(\phi,\pi)=f(Q,P)$ in terms of $Q,P$. The $Q,P$ are coordinates
on the reduced phase space which is eqipped with the pull -- back 
symplectic structure\footnote{This symplectic structure coincides with the 
pull -- back of the degenerate symplectic structure on the 
full phase space corresponding to the 
Dirac bracket induced by the system $\{S,F,G\}$ \cite{20}.} 
induced by the embedding of the constraint surface specified by $f$. 

The gauge fixing conditions also induce a reduced Hamiltonian $H_r$ which 
only depends on $Q,P$ and which arises by computing the equations of 
motion 
for $Q,P$ with respect to $H_c$ and then restricting them to the gauge 
fixed values of the Lagrange multipiers and to the constraint surface.
Then $H_r$ is defined as the function of $Q,P$ only\footnote{For 
simplicity, we are 
assuming a gauge fixing which leads to a conservative reduced 
Hamiltonian.} which generates these same equations of motion.
We are now in the situation of an ordinary Hamiltonian system equipped 
with a true Hamiltonian $H_r$. We quantise a suitable subalgebra of the
reduced Poisson algebra as a $\ast-$algebra $\mathfrak{A}$ and represent 
it on a Hilbert space $\cal H$. This Hilbert space is to be identified 
with the physical Hilbert space arising from reduced phase space 
quantisation. Its relation with Dirac's constraint quantisation is spelled out in
\cite{9b}. Let 
$t\mapsto U(t)$ be the unitary 
evolution induced by $H_r$. Then the object of interest is the 
transition 
amplitude or n-point function
\be \label{5.14}
<\psi_f,U(t_f-t_n) a_n U(t_n - t_{n-1}) a_{n-1} .. U(t_2-t_1) a_1 
U(t_1-t_i)\psi_i>
\ee
between initial and final states $\psi_i,\psi_f$ at initial 
and final times $t_i,t_f$ respectively with 
intermediate measurements of the operators $a_1,..,a_n\in \mathfrak{A}$ at 
$t_1<t_2<..<t_n$. 

Preferrably one would like to be in a situation in which there is a
cyclic vector $\Omega$ for $\mathfrak{A}$ which is also a ground state 
for $H_r$. The existence of a cyclic vector is 
no restriction because representations of $\mathfrak{A}$ are always 
direct 
sums of cyclic representations. In this case $\mathfrak{A}\Omega$ is dense 
in $\cal H$ and we may therefore restrict attention to 
$\psi_i=\psi_f=\Omega$ by choosing appropriate $a_1,..,a_n$ in 
(\ref{5.14}). The existence of a vacuum state for $H_r$ means that 
zero is in the point spectrum of $H_r$. Let us make this assumption for 
simplicity. 

Let us abbreviate the 
Heisenberg time evolution as
$a_k(t):=U(t)^{-1}a_k U(t)$. 
In principle it  
would be sufficient to restrict the $a_k$ to 
be configuration operators $Q$ because their time evolution
contains sufficient information about $P$ as well. However, we will
stick to the more general case for reasons that will become clear later.
This leads us the n-point function
\be \label{5.15}
S(t_1,..,t_n):=\frac{<\Omega,U(t_f) a_n(t_n).. a_1(t_1) U(-t_i)\Omega>} 
{<\Omega,U(t_f-t_i)\Omega>} 
\ee
where we have properly normalised as to give the 0 -- point function the 
value unity. This has the advantage that certain infinities that 
woudl otherwise arise 
in the following can be absorbed. Notice that since $\Omega$ is a 
ground state, the $U(t_f)$ and $U(t_i)$ as well as the 
denominator could be dropped in (\ref{5.15}).

Now a combination of well known heuristic arguments \cite{20}, \cite{24} 
reviewed 
in 
\cite{9b} reveals the following:\\
Consider any initial and final configuration $q_i,q_f$ on the full
phase space and denote by ${\cal P}((t_i,q_i),(t_f,q_f))$ the set 
of paths\footnote{This should be a suitable measurable space but 
we leave it unspecified.} in full configuration space between 
$(q_i,q_f)$ at times $t_i,t_f$ respectively. Consider
\be \label{5.25}
Z[j;q_i,q_f]=\lim_{-t_i,t_f\to \infty}
\int_{{\cal P}((t_i,q_i),(t_f,q_f))} \; 
[Dq\;Dp\;D\lambda\;D\mu]\;\delta[G]\;|\det[\{F,G\}]|\;\rho\;
e^{\frac{i}{\hbar} 
S[q,p,\lambda,\mu]}
\;e^{i\int_{t_i}^{t_f}\; dt\; j(t)\cdot q(t)}
\ee
Here $j$ is a current in the fibre bundle dual to that of $q$, 
$S[q,p,\lambda,\mu]$ is the canonical action after performing the 
singular Legendre transform from the Lagrangian to the Hamiltonian 
formulation\footnote{The Lagrange multipliers $\lambda,\mu$ of the 
primary 
first and 
second class constraints respectively play the role of the velocities 
that one could not solve for in terms of the momenta in the process of 
the Legendre transform.} and $\rho$ is a local function of $q,p$ which
is usually related to the Dirac bracket determinant $\det[\{S,S\}]$ 
\cite{24}.

Now the primary constraints are always of the
form $\pi=f(Q,P,\phi)$ where we have split again the canonical pairs into
two groups. Thus, $S[q,p,\lambda,\mu]$
is linear in those momenta $\pi$ and we can integrate them out yielding
$\delta$ distributions of the form
$\delta[\lambda - (.)]\; \delta[\mu - (.)]$ which can be solved by
integrating over $\lambda,\mu$. If we assume that the dependence
of the remaining action on $P$ is only quadratic and that $G$ and
$|\det[\{F,G\}]|$ are independent of $P$ then we can integrate also over
$P$ which yields in general a Jacobian $I$ coming from the Legendre
transform. We can then write (\ref{5.25}) as
\be \label{5.25a}
Z[j;q_i,q_f]=\lim_{-t_i,t_f\to \infty}
\int_{{\cal P}((t_i,c_i),(t_f,c_f))} \;
[Dq]\;\delta[G]\;|\det[\{F,G\}]|\;\rho\;I\;
e^{\frac{i}{\hbar}
S[q]}
\;e^{i\int_{t_i}^{t_f}\; dt\; j(t)\cdot q(t)}
\ee
where proper substitutions of $\pi$ from solving the primary constraints
and of $P$ from the Legendre transformation are understood. Here $S[q]$
is the original (covariant) Lagrangian action.

Defining $\chi[j]:=\frac{Z[j;\;q_i,q_f]}{Z[0;\;q_i,q_f]}$ the 
covariant or path integral n -- point 
functions 
\be \label{5.26}
S(t_1,..,t_n):=[\frac{\delta^n \chi[j]}{i^n \delta j(t_1) .. \delta 
j(t_n)}]_{j=0}
\ee
have the canonical or physical interpretation of 
\be \label{5.27}
<\Omega, T(a_1(t_1)..a_n(t_n)) \Omega>
\ee
where $T$ is the time ordering symbol, $\Omega$ is the aforementioned 
cyclic 
vacuum vector 
defined by the physical (or reduced) Hamiltonian $H_r$ induced by the 
gauge fixing $G$, $a_k(t)$ is the Heisenberg operator at time $t$
(evolved with respect to $H_r$)
corresponding to $a_k$ and $a_k$ itself classically corresponds to 
a component of $q$ 
evaluated on the constraint surface $S=F=G=0$. The scalar product 
corresponds to a quantisation on the reduced phase space defined 
by $G$. Notice how the gauge 
fixing condition $G$ (or choice of clocks) prominently finds its way 
both into the 
canonical theory and into the path integral formula (\ref{5.25a}). 
In particular, notice that the seemingly similar expression
\be \label{5.28}
Z'[j;q_i,q_f]=\lim_{-t_i,t_f\to \infty}
\int_{{\cal P}((t_i,q_i),(t_f,q_f))} \; 
[Dq]\;
e^{\frac{i}{\hbar} 
S[q]}
\;e^{i\int_{t_i}^{t_f}\; dt\; j(t)\cdot q(t)}
\ee
does not have any obvious physical interpretation and in addition lacks 
the important measure factors $\rho,\;I$. \\
\\
Remarks:
\begin{itemize}
\item[1.]
One may be puzzled by the following: From ordinary gauge theories on  
background spacetimes such as Yang -- Mills theory on Minkowski space 
the path integral or more precisely the generating functional of the 
Schwinger functions (in the Euclidian formulation) does not require 
any gauge fixing in order to give the path integral a physical 
interpretation. One needs it only in order to divide out the gauge 
volume in a systematic way (Fadeev -- Popov identity) while the 
generating functional  
is independent of the gauge fixing. The gauge fixing also does not 
enter the construction of gauge invariant functions (such as Wilson 
loops). In our case, however, the gauge 
fixing condition is actually needed in order to formulate the physical 
time evolution and the preferred choice of gauge invariant functions on 
phase space. 

The resolution is as follows: The difference between Yang -- Mills  
theory and generally covariant systems such as General Relativity 
that we are interested in here is indeed that the canonical 
Hamiltonian is in fact the generator of gauge 
transformations (spacetime diffeomorphisms) rather than physical time 
evolution. It is even constrained to vanish. In contrast, in Yang -- 
Mills theory there is a preferred and gauge invariant Hamiltonian 
which is not constrained to vanish. In order to equip the theory 
with a notion of time we have 
used the relational framework discovered in \cite{25} which consists in 
choosing fields as clocks and rods with respect to which other fields 
evolve. Mathematically this is equivalent to a choice of gauge fixing.
Hence, in our case the gauge fixing plays a dual role: First, in order
to render the generating functional less singular and secondly in order 
to define physical time evolution. 
\item[2.]
The appearance of the $\delta$ distributions and functional 
(Fadeev -- Popov) determinants 
in (\ref{5.25}) indicates that we are not dealing with an ordinary 
Hamiltonian system but rather with a constrained system. One can in fact 
get rid of the gauge fixing condition involved if one pays a price. The 
price is 
that if one considers instead of $q$ its gauge invariant extension
$\tilde{q}$ off the surface $G=0$ \cite{20,22}, then, since we 
consider
the quotient $Z[j]/Z[0]$ which leads to {\it connected n -- point 
functions}, by the usual Fadeev -- Popov identity that exploits 
gauge invariance we may
replace \cite{9b} (\ref{5.25}) by
\be \label{5.24}
\tilde{Z}[j,q_i,q_f]=
\int_{{\cal P}((t_i,q_i),(t_f,q_f))} \; [Dq\;Dp\;D\lambda\;D\mu]\;\rho 
e^{\frac{i}{\hbar} 
S[q,p,\lambda,\mu]}
\;e^{i\int_{t_i}^{t_f}\; dt\; j(t)\cdot \tilde{q}(t)}
\ee
However, (\ref{5.24}) is not very useful unless $\tilde{q}(q,p)$ is easy 
to calculate which is typically not the case. Hence, we will refrain 
from doing so. 	Nevertheless, no matter whether one deals with 
(\ref{5.25}) or (\ref{5.24}), the correlation functions depend on the 
gauge fixing $G$ or in other words on the choice of the clocks 
\cite{22,23} with respect to which one defines a physical reference 
system.
\item[3.] The correspondence between between (\ref{5.26}) and 
(\ref{5.27}) also allows to reconstruct the physical inner product
from the n -- point functions: given arbitrary states $\psi,\psi'\in 
{\cal H}$ 
we find $a,a'\in \mathfrak{A}$ such that 
$||a\Omega-\psi||,\;||a'\Omega-\psi'||$ are arbitrarily small. Now pick 
any $t_i<t_0<t_f$ then
\be \label{5.27a}
<a\Omega,a'\Omega>=<\Omega,a^\dagger\;a'\Omega>
\ee
By assumption, the operator $a^\dagger a'$ can be written as a finite 
linear combination of monomials of homogeneous degree in the 
components of the operator $q$ which we write, suppressing indices for 
the components as $q^n$. Then
\be \label{5.27b}
<\Omega, q^n \Omega>=\lim_{t_1,..,t_n\to t_0; 
t_n>..>t_1}\; <\Omega,q(t_n)..q(t_1)\Omega>
\ee
which can be expressed via (\ref{5.26}).
The existence of this coincidence limit of n -- point functions is 
often problematic in background dependent Wightman QFT \cite{Haag} but 
their existence is actually the starting point of canonical quantisation
of background independent non -- Wightman QFT as one can see from the 
identity (\ref{5.27b}).

\end{itemize}

\subsection{The Generating Functional of Tetrad n -- point functions}
\label{s5.3}

We now want to apply the general framework of the previous section to 
General Relativity in the Holst formulation. Classically its is clear
that without fermions all the geometry is encoded in the co-tetrad 
fields 
$e^I_\mu$ because then the spacetime connection is just the spin 
connection defined by the co-tetrad (on shell). If fermions are coupled, 
the same 
is still true in the second order formulation so that there is no 
torsion. But even in the first order formulation with torsion one 
can atttribute the torsion to the fermionic degrees of freedom. Hence 
we want to consider as a complete list of configuration fields the co -- 
tetrad. 

We will now make two assumptions about the choice of gauge fixing and 
the matter content of our system.
\begin{itemize}
\item[I.] The local measure factors $\rho,\;I$ depends on the co -- tetrad
only analytically. This is actually true for the Holst action
\cite{8a}, see also \cite{8c}. 
\item[II.] The gauge fixing condition $G$ is independent of the 
co -- tetrad and the Fadeev -- Popov 
determinant $\det(\{F,G\})$ depends only analytically on
the co -- tetrad. 
With respect to the first class Hamiltonian and spatial diffeomorphism 
constraint this 
can always be achieved by choosing suitable matter as a reference 
system, see e.g. \cite{26}. However, in addition there is the Gauss -- 
law first class constraint. Here it is customary to impose the time 
gauge gauge condition \cite{7} which asks that certain components of the 
tetrad vanish. This will also enable one to make the connection with 
canonical LQG where one works in the time gauge in order to arrive at an
SU(2) rather than G connection. Fortunately, in this case it is 
possible to explicitly construct a complete set of G -- invariant 
functions of the tetrad, namely the four metric\footnote{In the presence 
of fermions there are additinal gauge invariant functions also 
involving the fermions.}  $g_{\mu\nu}=e^I_\mu 
e^J_\nu \eta_{IJ}$ and if we only consider correlators of those then we 
can get rid of the time gauge condition as indicated in the previous 
section (Fadeev -- Popov identity). In section \ref{s5.5} we will come 
back to this issue, however, in trying to make the connection of the 
SFM obtained with canonical LQG for which the time gauge is unavoidable.
We will then sketch how to possibly relax the assumptions made under 
[II.].
\end{itemize}
Under the assumptions made we consider the generating functional
$\chi(j,J):=Z[j,J]/Z[0,0]$ where 
\ba \label{5.29}
Z[j,J] &:=& \int [D\phi\;\;DA]\; \delta[G[A,\phi]]\; 
e^{i\int_M\; {\rm Tr}(J\wedge \phi)}\;
\times\nonumber\\
&& \int\; [De]\; \rho[e,A,\phi]\; I[e,A,\phi]\; 
|\det[\{F,G\}]|[e,A,\phi]
\; e^{\frac{i}{\hbar} 
(S_g[e,A]+S_m[e,A,\phi])}
e^{i\int_M\; {\rm Tr}(j\wedge e)}
\ea
Here $\phi$ denotes the matter configuration variable.
We have split the 
total action into the geometry 
(Holst) part $S_g$ and a matter part $S_m$ which typically depends non 
trivially but analytically on $e$. Also the total current was split into 
pieces $J,j$ 
respectively taking values in the bundles dual to those of $\phi,e$ 
respectively. 

A confusing and peculiar feature of first order actions
such as the Holst or Palatini action is that from a Lagrangian point of 
view both fields $e,A$ must be considered as configuration variables.
In performing the Legendre transform \cite{8a}
one discovers that there are 
primary 
constraints which relate certain combinations of $e$ to the momenta 
conjugate to $A$. One can solve these constraints and then $(A,e)$ appear 
as momentum and configuration coordinates of this partly reduced phase 
space. This is the reason why 
consider only correlations with respect to $e$.   

The idea is now as usual in path integral theory: We set 
\be \label{5.30}
\sigma[e,A,\phi]:=\rho[e,A,\phi]\; I[e,A,\phi]\; 
|\det[\{F,G\}]|[e,A,\phi]
\; e^{\frac{i}{\hbar} S_m[e,A,\phi]}
\ee
and write (\ref{5.29}) as
\be \label{5.31}
Z[j,J]:=\int [D\phi\;DA]\; \delta[G[A,\phi]]\; 
e^{i\int_M\; {\rm Tr}(J\wedge \phi)}\; 
\sigma[\frac{\delta}{i\delta j},A,\phi]\;
\{
\int\; [De]\;  e^{\frac{i}{\hbar} 
S_g[e,A]}
e^{i\int_M\; {\rm Tr}(j\wedge e)}\}
\ee
Of course $e^{iS_m}$ must be power expanded in a perturbation series in 
order to carry out the functional derivations with resepect to $j$. 
Indeed, if 
we consider just the functional integration with respect to $e$ and 
think of $A,\phi$ as external fields then the piece $S_g$ being 
quadratic in $e$ is like a {\it free part} while $S_m$ being only 
analytic in 
$e$ is like an {\it interaction part} of the action as far as the 
co-tetrad is 
concerned. Of course, in the computation of the physical tetrad n -- 
point functions all the functional derivatives involved in (\ref{5.31}) 
are eventually evaluated at $j=0$.   

It follows that the object of ultimate interest is the {\it Gaussian 
integral}
\be \label{5.32}
z[j;A]:=\int\; [De]\;  e^{-\frac{i}{\ell_P^2} \;\int_M\; {\rm 
Tr}(G\wedge 
e\wedge e+\ell_P^2 j\wedge e)}
\ee
which is computable exactly. Of course, it is not 
a standard Gaussian, first because the exponent is purely imaginary. 
Secondly because the ``metric'' $G^{\mu\nu}_{IJ}(A)$ is indefinite so 
that
$z[j;A]$ would  be ill defined if the exponent was real\footnote{As 
usual
this prevents a ``Euclidian'' version of GR. Here Euclidian stands 
for Euclidian field theory with an analytic continuation to the 
imaginary axis of the real time variable involved (Wick rotation) which 
leads to a real exponent. 
This has nothing to do with Lorentzian or Euclidian signature GR.
In fact, most metrics do not have an analytic section so that Wick 
rotation is ill defined and thus the connection between the real and the 
Euclidian theory is veiled.}. 
In the 
appendix we remind the reader how to integrate such non -- standard 
Gaussians. In order to carry out this integral we must make the 
technical assumption that configurations $A$ for which $G$ is singular
have measure zero with respect to $DA$. We will come back to this 
assumption later. 

It is at this point where we must regularise the path integral in order 
to perform the Gaussian integration\footnote{Actually we can formally solve 
the Gaussian integral w/o specifying
the triangulation, i.e. we can compute it in the continuum. However, 
then one must regularise the resulting determinant which amounts to the 
same problem.} and we write the discretised version 
on a cubulation of $M$ as motivated in section \ref{s5.1}, that is, we 
replace (\ref{5.32}) by the discretised version
\be \label{5.33a}
z[j;A]:=\int\; \prod_{v,I,\mu}\; de^I_\mu(v)\;  e^{\frac{i}{\ell_P^2} 
\;\sum_v\;[G^{\mu\nu}_{IJ}(v) e^I_\mu(v) e^J_\nu(v)+\ell_P^2 j^I_\mu(v)
e^I_\mu(v)]}
\ee
The results of appendices \ref{sa} and \ref{sb} now reveal that 
\be \label{5.33}
z[j;A]:=[\prod_{v}\; \frac{e^{\frac{i\pi}{4} {\rm 
ind}(G(v))}}{\sqrt{|\det(G(v))|}}] \; e^{-i\frac{\ell_P^2}{4}\sum_v 
[G^{-1}(v)]_{\mu\nu}^{IJ} j^\mu_I(v) j^\nu_J(v)}
\ee
where we dropped a factor $\sqrt{\pi}^{16N}$ for a cubulation with
$N$ vertices because it is cancelled by the same factor coming from the 
denominator in $\chi(j,J)$, see (\ref{5.29}).

\subsection{Wick Structure, Graviton Propagator and SFM Vertex 
Structure}
\label{s5.4}

\subsubsection{Wick structure}
\label{s5.4.1}

Formula (\ref{5.33}) explicitly displays the main lesson of our 
investigation: The full $j$ dependence of the generating functional 
written as (\ref{5.31}) rests in (\ref{5.33}). We are interested 
in the n-th functional derivatives of (\ref{5.33}) at $j=0$. Now similar 
as 
in free field theories, the corresponding n -- point functions 
vanish for $n$ odd. However, in contrast to free field theories,
for $n$ even, the $n-$point functions cannot be written in terms of 
polynomials of the 2-point function. The reason is that the 
``covariance'' $G^{-1}[A]$ of the Gaussian is not a background structure
but rather depends on the quantum field $A$ itself that one has to 
integrate over. 
This renders the co -- tetrad theory to be a non -- quasi -- free, that 
is,
interacting theory. Nevertheless it is true that all Wick identities 
that have been derived for free field theories still hold also for 
the $n-$point tetrad functions {\it albeit in the sense of
expectation values or means with respect to $A$}. 

\subsubsection{Graviton Propagator}
\label{s5.4.2}

To illustrate this, consider a fictive theory in which 
$\sigma(e,A,\phi),\;G(A,\phi)$ are both independent of 
both $A,e$. This is not a very physical assumption but it serves
to make some observations of general validity in a simplified context.
This means that we can drop the $\phi$ dependence because the 
generating functional factorises.
Thus in our fictive theory we are looking at the generating functional
$\chi[j]=z[j]/z[0]$ where 
\be \label{5.34}
z[j]=\int\; [DA]\; z[j;A]=\int\; \prod_{v,\mu} \; d\mu_H(A(l_\mu(v))\; 
[\prod_v \frac{e^{\frac{i\pi}{4} {\rm 
ind}(G(v))}}{\sqrt{|\det(G(v))|}}]\;
e^{-\frac{i\ell_P^2}{4}\sum_v \; j^\mu_I(v)\; j^\nu_J(v)\; 
[G^{-1}(v)]^{IJ}_{\mu\nu}]}
\ee
where $\mu_H$ is the\footnote{In case of non -- compact $G$ the Haar 
measure is unique up to a normalisation constant which drops out in 
$\chi(j)$. To choose the Haar measure instead of the Lebesgue measure 
makes sense in the continuum limit of infinitely ``short'' edges as usual.} 
Haar measure 
on $G$. Now let 
\be \label{5.35}
<e_{\mu_1}^{I_1}(v_1)\; ..\; e_{\mu_n}^{I_n}(v_n)>
:=[\frac{\delta^n \chi[j]}{i^n 
\delta j^{\mu_1}_{I_1}(v_1)\; .. \; \delta j^{\mu_1}_{I_1}(v_1)}]_{j=0}
\ee
It is immediately clear that 
\be \label{5.36}
<e_{\mu_1}^{I_1}(v_1)\; e_{\mu_2}^{I_2}(v_2)>=0
\ee
unless $v_1=v_2$. This is reassuring because as we said above, 
physically it makes only sense to consider correlators of $G-$invariant 
objects such as the metric. The simplest $n-$point function of interest 
is therefore the 4-point function
\be \label{5.37}
< g_{\mu_1 \nu_1}(v_1) \; g_{\mu_2 \nu_2}(v_2)> 
=  
<e_{\mu_1}^{I_1}(v_1)\; e_{\nu_1}^{J_1}(v_1)
e_{\mu_2}^{I_2}(v_2)\; e_{\nu_2}^{J_2}(v_2)>\; \eta_{I_1 J_2}\; 
\eta_{I_2 J_2}
\ee
If we are interested in something like a graviton propagator we are 
interested in $v_1\not=v_2$ and obtain 
\be \label{5.38}
< g_{\mu_1 \nu_1}(v_1) \; g_{\mu_2 \nu_2}(v_2)> 
=[\frac{\ell_P^2}{2}]^4 
<[G(v_1)^{-1}]^{I_1 J_1}_{\mu_1 \nu_1}\; 
[G(v_2)^{-1}]^{I_2 J_2}_{\mu_2 \nu_2}>'
\ee
where for $F=F[A]$
\be \label{5.39}
<(F)>':=
\frac{
\int\; \prod_{v,\mu} \; d\mu_H(A(l_\mu(v))\; 
[\prod_v \frac{e^{\frac{i\pi}{4} {\rm 
ind}(G(v))}}{\sqrt{|\det(G(v))|}}]\;F[A]}
{\int\; \prod_{v,\mu} \; d\mu_H(A(l_\mu(v))\; 
[\prod_v \frac{e^{\frac{i\pi}{4} {\rm 
ind}(G(v))}}{\sqrt{|\det(G(v))|}}]
}
\ee
Notice that $G(v)^{-1}$ does not share the symmetries of $G(v)$, so 
$[G^{-1}(v)]^{(IJ)}_{\mu\nu}$ does not vanish automatically.
We see that we are basically interested in correlators of the inverse 
matrix $G(v)^{-1}$ with respect to the joint Haar measure.  
Whether these have the correct behaviour in a situation where,  
instead of vacuum boundary states, one chooses coherent states peaked 
on a classical background metric as suggested in \cite{6g,9a} is 
currently under investigation.

\subsubsection{Spin Foam Vertex Structure}
\label{s5.4.3}

Finally, in order to translate (\ref{5.39}) into spin foam language,
we should perform harmonic analysis on $G$ and write the integrand of 
the Haar measure in terms of irreducible representations of $G$. 
In particular, the vertex structure of a spin foam is encoded in $z[0]$
so that we are interested in harmonic analysis of the function 
\be \label{5.40}
F(v):=\frac{e^{\frac{i\pi}{4} {\rm ind}(G(v))}}{\sqrt{\det(G(v))|}}
\ee
To derive its a graph theoretical structure it is enough to find out 
which $F(v)$ depend how on a given holonomy $A(l)$. Recall that $F(v)$ 
is 
a function cylindrical over the graph 
$\gamma(v)=\cup_{\mu<\nu} \partial f_{\mu\nu}(v)$ which is the union of 
its respective plaquette loops . Consider a fixed 
edge $l=l_\mu(v)$. It is contained in $\gamma(v')$ if and only if 
it is contained in one of the plaquette loops $\partial f_{\mu \nu}(v')$ 
or $\partial f_{\nu \mu}(v')$ with $\mu<\nu$ or $\nu<\mu$ respectively.
In the first case it must coincide either with $l_\mu(v')$ or with 
$l_\mu(v'+\hat{\nu})$. In the second case it must coincide either with 
$l_\mu(v'+\hat{\nu})$ or with $l_\mu(v')$ as well. Thus in either case we 
must have either $v'=v$ or $v'=v-\hat{\nu},\;\nu\not=\mu$. 

For our illustrative purposes let us consider for simplicity that $G$ 
is compact, the non compact case has the same spin foam vertex structure but 
the harmonic analysis is a bit more complicated. Then each function 
$F(v)$ can be formally expanded into $SO(4)$ (or rather the 
universal cover $SU(2)\times 
SU(2)$) irreducible 
representations\footnote{This expansion would be rigorous if we 
knew that $F(v)$ is an $L_2$ function which is currently under 
investigation. We assume here that in any case we  may use the 
Peter \& 
Weyl theorem in a distributional sense.} with respect to the 
six plaquette holonomies $A(\partial f_{\mu\nu}(v)),\;\mu<\nu$. 
These representations $\pi$ are labelled by pairs of half integral spin 
quantum numbers but we will not need this for what follows.
Thus $F(v)$ admits an expansion of the form
\be \label{5.41}
F(v)=\sum_{\{\pi_{\mu\nu}\}}\; \iota'_{\{\pi_{\mu\nu}\}}\; \cdot
\;[\otimes_{\mu<\nu} \; \pi_{\mu\nu}(A(\partial f_{\mu\nu}(v)))]
\ee
where $\iota'_{\{\pi_{\mu\nu}\}}$ is a gauge invariant intertwiner
for the six -- tuple of irreducible representations 
$\{\pi_{\mu\nu}\}_{\mu<\nu}$ which is independent of $v$, the only $v$ 
dependence rests in the holonomies. It depends on the specific algebraic 
form of $F(v)$ which derives from the Holst action.

Let us define $\pi_{\nu\mu}:=\pi_{\mu\nu}$ for $\mu<\nu$. By writing 
the six plaquette holonomies in terms of four edge holonomies it is not 
difficult to 
see that $F(v)$ can also be written in the form  
\be \label{5.42}
F(v)=\sum_{\{\pi_{\mu\nu}\}}\; \iota_{\{\pi_{\mu\nu}\}}\;\cdot\;
[\otimes_{\mu,\mu\not=\nu} \; \pi_{\mu\nu}(A(l_\mu(v)))\;\otimes \;
\pi_{\mu\nu}A(l_\mu(v+\hat{\nu}))]
\ee
which displays explicitly the 16 
variables $A(l_\mu(v)),\;A(l_\mu(v+\hat{\nu}),\;\nu\not=\mu$ involved 
and 
consists of 24=6 x 4 tensor product factors. In order to arrive at 
(\ref{5.42}) we had to rearrange the contraction indices which 
induces the change from $\iota'$ to $\iota$ and we have 
made use of $\pi(A(l)^{-1})=\pi^T(A(l))$ for $G=SO(4)$. 

We may now carry out explicitly the integrals over edge holonomies in 
$z[0]$ by inserting the expansion (\ref{5.42}). We write 
symbolically\footnote{We rearrange the tensor products as if they were 
scalars but this can be corrected by performing corresponding 
rearrangements in the contraction structure of the intertwiners. We
assume this to be done without explicitly keeping track of it because 
it does not change the vertex structure.}
\ba \label{5.43}
z[0] &=& \int \; \prod_{v,\mu} \; d\mu_H(A(l_\mu(v)))\;\prod_{v'}\; 
F(v')
\nonumber\\
&=& \sum_{\{\pi^v_{\mu\nu}\}}\; [\prod_v \; 
\iota_{\{\pi^v_{\mu\nu}\}} \cdot]
\int \; \prod_{v,\mu} \; d\mu_H(A(l_\mu(v)))\;
[\otimes_{v',\mu,\mu\not=\nu} \; 
\pi^{v'}_{\mu\nu}(A(l_\mu(v')))\;\otimes 
\;
\pi^{v'}_{\mu\nu}(A(l_\mu(v'+\hat{\nu}))]
\nonumber\\
&=& \sum_{\{\pi^v_{\mu\nu}\}}\; [\prod_v \; 
\iota_{\{\pi^v_{\mu\nu}\}} \cdot]
\; \int \; \prod_{v,\mu} \; d\mu_H(A(l_\mu(v)))\;
[\otimes_{v',\mu,\mu\not=\nu} \; 
\pi^{v'}_{\mu\nu}(A(l_\mu(v')))\;\otimes 
\;
\pi^{v'-\hat{\nu}}_{\mu\nu}(A(l_\mu(v'))]
\nonumber\\
&=& \sum_{\{\pi^v_{\mu\nu}\}}\; [\prod_v \; 
\iota_{\{\pi^v_{\mu\nu}\}} \cdot]
\; \otimes_{v,\mu} [\int_G\; d\mu_H(g)\;
[\otimes_{\mu\not=\nu} \; 
\pi^{v}_{\mu\nu}(g)\;\otimes 
\;
\pi^{v-\hat{\nu}}_{\mu\nu}(g)]
\ea
Here in te second step we have shifted the vertex label in one of the 
tensor product factors in order to bring out the dependence on the 
$A(l_\mu(v))$. It follows that the end result of the integration is that 
for each edge $l=l_\mu(v)$ there is a gauge invariant 
intertwiner\footnote{We do not bother to expand it into a recoupling 
scheme and thus to label the intertwiner itself by three irreducible 
representations.}
\be \label{5.44}
\rho_{\{\pi^v_{\mu\nu},\pi^{v-\hat{\nu}}_{\mu\nu}\}_{\nu\not=\mu}}
:=
[\int_G\; d\mu_H(g)\;
[\otimes_{\mu\not=\nu} \; 
\pi^{v}_{\mu\nu}(g)\;\otimes 
\;
\pi^{v-\hat{\nu}}_{\mu\nu}(g)]
\ee
which intertwines {\it six} representations rather than four as in 
(constrained) BF theory on simplicial triangulations. The origin of this 
discrepancy is of course that we are using cubulations rather than 
simplicial triangulations. These six representations involved for 
edge $l_\mu(v)$ correspond precisely to the six plaquette loops 
$\partial f_{\mu\nu}(v),\;\partial 
f_{\mu\nu}(v-\hat{\nu}),\;\nu\not=\mu$
of which $l_\mu(v)$ is a segment. Therefore, if we associate to each 
face $f=f_{\mu\nu}(v)$ an irreducible representation 
$\pi_f=\pi^v_{\mu\nu}$ and denote by $\{\pi\}$ the collection of all the 
$\pi_f$ 
then the  
basic building block
(\ref{5.44}) can be written in the more compact form 
\be \label{5.45}
\rho_l[\{\pi\}]=\int_G\; d\mu_H(g)\; \otimes_{l\subset \partial f}\;
\pi_f(g)
\ee
Likewise, if we denote $\iota_v[\{\pi\}]:=\iota_{\{\pi^v_{\mu\nu}\}}$,
then
\be \label{5.46}
z[0]=\sum_{\{\pi\}}\; [\prod_v\; \iota_v[\{\pi\}]\cdot]\;
\otimes_l \; \rho_l[\{\pi\}]
\ee
which of course hides the precise tensor product and contraction 
structure but is sufficient for our purposes. 

\begin{figure}[hbt]
\begin{center}
\psfrag{0+}{$l^+_0$}
\psfrag{1+}{$l^+_1$}
\psfrag{2+}{$l^+_2$}
\psfrag{3+}{$l^+_3$}
\psfrag{0-}{$l^-_0$}
\psfrag{1-}{$l^-_1$}
\psfrag{2-}{$l^-_2$}
\psfrag{3-}{$l^-_3$}
\includegraphics[scale=0.5]{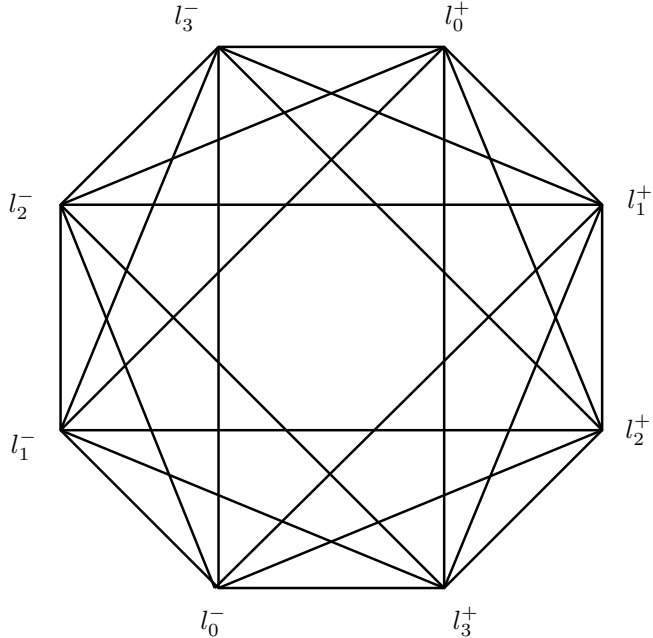}
\caption{The octagon diagramme associated to vertex $v$. The eight 
corners correspond to the eight edges 
$l=l_\mu^\sigma(v)=l_\mu(v+\frac{\sigma-1}{2}\hat{\mu}),\;\sigma=\pm$
adjacent to $v$. The line between
corners labelled by $l^\sigma_\mu(v),\;l^{\sigma'}_\nu(v)$ 
for $\mu\not=\nu$ corresponds 
to the face $f=f_{\mu\nu}^{\sigma\sigma'}(v)=
f_{\mu\nu}(v+\frac{\sigma-1}{2}\hat{\mu}+\frac{\sigma'-1}{2}\hat{\nu})$.
We should colour corners by intertwiners $\rho_l$ and lines by 
representations $\pi_f$ but refrain from doing so in order not to
clutter the diagramme. Altogether 48 irreducible 
representations of $Spin(4)$ (or 96 of $SU(2)$) are involved.}
\label{fig1}
\end{center}
\end{figure}

Formula (\ref{5.46}) is precisely the general structure of a SFM. 
Moreover, the intertwiner (\ref{5.45}) is the direct analog of the 
intertwiner in BF theory which there defines the pentagon diagramme
\cite{5}. If we would try to draw a corresponding picture for our 
model then for each vertex $v$ we would draw eight points, one for each  
each edge $l$ incident at $v$. These edges are labelled by the 
intertwiner $\rho_l$. Given two points corresponding to edges 
$l,l'$ consider the unique face $f$ that has $l,l'$ in its 
boundary. Draw a line between each such points and label it by $\pi_f$.
The result is the {\it octagon diagramme}, see figure \ref{fig1}. 
Concretely, the edges 
adjacent to $v$ are $l_\mu(v),\;l_\mu(v-\hat{\mu}),\;\mu=0,1,2,3$. For 
$\mu\not=\nu$,  
the face spanned by $l_\mu(v),\;l_\nu(v)$ is $f_{\mu\nu}(v)$, 
the face spanned by $l_\mu(v),\;l_\nu(v-\hat{\nu})$ is 
$f_{\mu\nu}(v-\hat{\nu})$,
the face spanned by $l_\mu(v-\hat{\mu}),\;l_\nu(v)$ is 
$f_{\mu\nu}(v-\hat{\mu})$ 
and finally
the face spanned by $l_\mu(v-\hat{\mu}),\;l_\nu(v-\hat{\nu})$ is 
$f_{\mu\nu}(v-\hat{\mu}-\hat{\nu})$. The corresponding label on the 
lines is thus $\pi_{\mu\nu}^v,\; \pi_{\mu\nu}^{v-\hat{\nu}},\; 
\pi_{\mu\nu}^{v-\hat{\mu}},\; 
\pi_{\mu\nu}^{v-\hat{\mu}-\hat{\nu}}$ respectively. Thus the octagon 
diagramme has eight points and 6 x 4 = 24 lines (each line connects two 
points). These correspond to the 
24 plaquettes that have a corner in $v$ namely for each 
$\mu<\nu$ these are $f_{\mu\nu}(v),\; f_{\mu\nu}(v-\hat{\mu}),\;   
f_{\mu\nu}(v-\hat{\nu}),\;f_{\mu\nu}(v-\hat{\mu}-\hat{\mu})$. In the 
case of $G=SO(4)$ each irreducible representation is labelled
by two spin quantum numbers. The intertwiner freedom is 
labelled by three irreducible representations of of $SO(4)$ and there is 
one irreducible representation corresponding to a face. Thus the octagon 
diagramme depends on 3 x 8 + 24=48 irreps of $SO(4)$ or 96 spin quantum 
numbers. Since each intertwiner (\ref{5.45}) factorises into {\it two} 
intertwiners \cite{3} (one for the starting point and one for the 
beginning 
point of the edge but both depending on the same representations) we 
may actually collect those eight intertwiners associated to the same 
vertex. The collection of those eight factors is 
actually the analytic expression corresponding to the octagon diagramme
which therefore may be called the {\it 96 j 
-- symbol}.
The decisive difference between (constrained) BF theory and our model is 
however that in (constrained) BF theory the analog of the function 
$F(v)$ is a product of $\delta$ distributions, one for each face 
holonomy. The simplicity constraints just impose restrictions on the
representations and intertwiners, but this cannot change the fact that
there is factorisation in the face dependence. In our model, the face 
dependence does not factorise, hence, in this sense it is less local or 
more interacting. 

To sum up this section:  we have explicitly described the analytical expression for the vertex amplitude of this SFM in (\ref{5.43}-\ref{5.46}), which
leads to the octagon diagramme described. Using harmonic analysis on $SO(4)=SU(2) \times SU(2)/\mathbb{Z}_2$ one can easily describe 
everything in terms of spin representations of SU(2). As the resulting expression is not very illuminating, we refrain from displaying it here.

\subsection{Relation between covariant and canonical connection}
\label{s5.5}

Another striking feature of our model is the following: Constrained 
BF theory, that is, Plebanski theory, should be a candidate for quantum 
gravity. Our Holst model should be equivalent to that theory at least 
semiclassically because morally speaking, the only difference 
between them lies in the technical implementation of the simplicity 
constraints, modulo the caveats mentioned in section \ref{s2}. Now 
one of the most important property of the implementation of the 
simplicity constraints in some of the most popular spin foam models is that the irreducible Spin(4) 
representations that one sums over are the simple ones\footnote{If we 
label an irrep of Spin(4) by a pair $(j_+,j_-)$ then a simple irrep. is 
one for which $j_+=j_-$ \cite{5}. There is a similar restriction if one 
works with arbitrary Immirzi parameter \cite{13}.}. In our model we do 
not see any sign of that\footnote{Also the models proposed in \cite{EteraMaite, 10, DanieleAristide} do not restrict to simple representations. 
The amplitudes are however peaked on these, though with a non-trivial width.}.
This is an important issue because the 
restriction to simple representations means that the underlying 
gauge theory is roughly SU(2) rather than Spin(4) which looks 
correct if the SFM is to arise from canonical LQG which indeed is a 
SU(2) gauge theory. Thus, in usual SFM the simplicity constraints 
seem to already imply the gauge fixing of the ``boost'' part of the 
Spin(4) 
Gauss constraint that is necessary at the classical level in order
to pass from the Holst connection to the Ashtekar -- Barbero -- Immirzi
connection \cite{7}. Strictly speaking, that has not been established 
yet as pointed out in \cite{26a} where it is shown that the connection 
used in SFM is actually the spin connection and not the Holst 
connection. But apart from that, in the considerations of the previous 
section we do not see any simplicity restrictions on the type of group 
representations.

However, notice that what we did in the previous section was incomplete
because in order to properly define the n -- point functions we must 
gauge fix the generating functional with respect to the G Gauss 
constraints. Formally this is not necessary if we only consider 
correlators of G invariant functions such as the metric because the 
infinite gauge group volume formally cancels out in the fraction 
$z[j]/z[0]$. However, details matter:\\ 
The formal arguments cannot be substantiated by hard proofs in this 
case. Specifically, if we consider $G=SO(1,3)$, there is no measure  
known for gauge theories for non compact groups (see \cite{26b} for the 
occurring complications) and 
thus 
we are forced to gauge fix at least the boost part of the Gauss 
constraint. This is the same reason for which one uses the time gauge
in the canonical theory. We expect 
that implementing the time gauge fixing \cite{7} in a way similar to the 
implementation of the simplicity constraints in usual BF theory will 
effectively reduce the gauge group to $SU(2)$. 

Details will appear 
elsewhere, but roughly speaking the idea is the following:\\
The time gauge is a set of constraints
$C[e]$ on the co -- tetrad $e$. By the usual manipulations we can 
pull the corresponding $\delta$ distribution out of the cotetrad 
fuctional integral and formally obtain  
\be \label{5.47}
\chi_{{\rm ABI}}[j]=[\delta[C[\delta/\delta j]]\; \chi_{{\rm Holst}}[j]
\ee
where $\chi_{{\rm Holst}}$ is the generating functional of the previous 
section and $\chi_{{\rm ABI}}$ stands for the Ashtekar -- Barbero -- 
Immirzi path integral. 

Whether this really works in a rigorous fashion remains to be seen. 
However, we find it puzzling that the simplicity constraints in usual
SFM, which classically have nothing to do with the time gauge, should 
automatically yield the correct boundary Hilbert space. It seems 
intuitively clear that the time gauge must be imposed in the quantum 
theory in addition to the simplicity constraints, just like in the 
classical theory, as we suggest. Without imposing it, we do not see 
any sign of a restriction from $G$ to $SU(2)$ in our model where 
we solve the simplicity constraints 
differently. We expect this to be related to the work \cite{26c}.
This observation indicates that usual SFM and our 
formulation are rather different from each other.

\section{Conclusions and Future Work}
\label{s6}

In this paper we have proposed a different strategy to construct spin 
foam models for GR. Rather than the Plebanski action we take the 
Holst action as our starting point. This means that the simplicity 
constraints of the Plebanski formulation have been correctly taken care 
of. The price to pay is that the connection to BF theory is lost.
The motivation behind our strategy is that BF theory is a TQFT and 
therefore quite different from GR which has an infinite number of 
physical degrees of freedom. Hence the usual triangulation independent 
methods developed for TQFT and employed in current SFM are possibly
less powerful in the context of GR. In particular, the fact that it 
is difficult to deal with the simplicity constraints in current SFM 
might be a sign of that. Another problem with the Plebanski formulation
that we have not mentioned yet is that it is difficult to couple 
matter because matter directly 
couples to the cotetrad rather than the B field. In principle one can 
express e via B modulo simplicity constraints but the corresponding 
formulas are even more involved than those for e. Notice that one 
{\it must} couple matter to BF theory in order to get a realistic 
model. For 3D
gravity coupling matter is straightforward \cite{FreidelLouapre} because there B 
field 
and e 
coincide
while in 4D this has not been done yet except for non standard model 
fermions which just couple to the connection \cite{Mikovic, Han} or 
membranes coupled to pure BF theory \cite{Perez}.

The method we proposed in this paper might be called a brute force 
and textbook strategy. Dropping the insistance on triangulation 
independence right from the beginning we proposed a Wilson action-like 
naive discretisation of the Holst action. We carefully studied the 
connection of the Holst path integral with the canonical LQG correlation 
or n -- point functions and used relational techniques to make the 
connection with the physical Hilbert space and observables. In 
principle, none of these ingredients are 
new, they have been successfully employed in other contexts. Of course,
the appealing elegance of (constrained) BF theory has disappeared in our 
formulation, the integrals to be computed are rather challenging (but 
not impossible) and the gauge fixing conditions for spatial 
diffeomorphism and Hamiltonian constraint as well as a local measure 
factor without which the connection to observation is lost complicate 
the formalism. Yet, we feel that the 
resulting structure, while far 
from being worked out in detail, has some interesting features such as 
the Wick like structure of the physical tetrad correlators and a less 
local structure. In particular, we have shown that imposing the time 
gauge also in the quantum theory comes out as a necessary and natural 
condition in our model in order to make contact with the LQG Hilbert 
space. 

As already mentioned in the introduction, this paper is exploratory in 
nature. It focuses more on ideas rather than analysis and there are many
open issues that need to be settled before the present model can be 
taken
seriously. Apart from the topological issues mentioned in section 
\ref{s5.1}, the convergence and measure theoretic issues discussed in 
section \ref{s5.3} and finally the issues with the imposition of the 
time gauge outlined in section \ref{s5.5} there are further points that 
need to be addressed. 

One of the most serious ones is the continuum limit: The fact that we are working with cubulations suggests a naive 
but natural notion of continuum limit which consists in studying the 
behaviour of the correlation functions under barycentric refinement of
the hypercubes at fixed IR regulator (boundary surface). The last couple of years has seen the resurgence of a coarse graining program  in spin foam models (see for e.g \cite{Bianca});
and it would be very interesting to see how the application of the procedure to our discretization compares with the standard models.  
Of course, in the spirit of the AQG framework \cite{AQG} one could also say
that the continuum limit has been taken already provided that one 
works with infinite cubulations. This requires then, in a separate
step, to remove the IR regulator. 

A more practical but still 
important problem is the following: Even at 
finite 
UV and IR regulator, it is already hard enough to compute the 
determinant of the covariance matrix of the co-tetrad Gaussian 
and to determine its index (which may vanish automatically, see appendix
\ref{sb}). But since these covariances are highly correlated, the 
practical computation of the n -- point functions at least in the 
macroscopic regime will be possible only 
if the corresponding non trivial measure has some kind of cluster 
property \cite{26d}. 

All of these issues are left for  future work.\\
\\
\\
\\
{\large Acknowledgements}\\
\\
We thank Bianca Dittrich and Kristina Giesel for illuminating 
discussions. We would also like to thank John Barrett for very useful discussions and comments.
The part of the research performed at the Perimeter Institute for 
Theoretical Physics was supported in part by funds from the Government of  
Canada through NSERC and from the Province of Ontario through MEDT.
This research project was supported in part by the European Science Foundation (ESF)
%by DOE-Grant
%DE-FG02-94ER25228 to Harvard University.

\begin{appendix}

\section{Non -- standard Gaussian integrals}
\label{sa}

Let $G$ be a real valued, symmetric and non singular matrix on the real 
vector space 
$V=\mathbb{R}^n$ and let $j\in V$. We are interested in the non -- 
standard Gaussian integral
\be \label{a.1}
I:=\int_V\;d^n x\; e^{i[\frac{1}{2}\; x^T G x+j^T x]}
\ee 
Using the translation $x=y-G^{-1} j$ we can simplify this to
\be \label{a.2}
I=e^{-i\frac{i}{2} j^T G^{-1} j}\; \int_V\;d^n y\; e^{\frac{i}{2}\; 
y^T G y}
\ee 
There exists an element $S\in GL(n,\mathbb{R})$ such 
that $G=S^T\; D\;S$ where $D$ is a regular diagonal n x n matrix
which is possibly indefinite.   
Denote by $d_1,..,d_n\;\in\;\mathbb{R}-\{0\}$ the entries of $D$. 
Then the change of coordinates $y=S z$ reveals 
\be \label{a.3}
I=\frac{e^{-i\frac{i}{2} j^T G^{-1} j}}{|\det(S)|}\; \int_V\;d^n z\; 
e^{\frac{i}{2}\; 
z^T D z}
\ee 
Now consider the basic integral
\be \label{a.4}
I_d:=\int_{\mathbb{R}}\; dz \; e^{id z^2/2}
\ee
for $d\in \mathbb{R}-\{0\}$. For $d=ik,\; k>0$ we know the value of  
(\ref{a.4}), however, that formula involves a square root and thus 
analytic continuation of $I_d$ in $d$ is ambiguous. Hence we must 
determine the 
value 
of (\ref{a.4}) by independent means.

The integrand in (\ref{a.4}) is entire analytic in $z$ 
without 
poles. For $d>0$ or $d<0$ respectively the 
integral over the arc $0\le {\rm arg}(z) \le \pi/2$ or
$0\ge {\rm arg}(z) \ge -\pi/2$ respectively vanishes at 
infinite radius. Hence, using a Cauchy integral argument we
may rotate the integral from $z\in \mathbb{R}$ to $z\in e^{i{\rm sgn}(d) 
\pi/4} \mathbb{R}$ so that with $z=e^{i{\rm sgn}(d)\pi/4}t,\;t\in 
\mathbb{R}$ we get 
\be \label{a.5}
I_d=e^{i{\rm sgn}(d)\pi/4}\;\int_{\mathbb{R}}\; dt \;e^{-|d| t^2/2}
=\sqrt{2\pi/|d|}\; e^{i{\rm sgn}(d)\pi/4} 
\ee
Given a symmetric matrix $G$ with signature $p,q$ (i.e. $p$ positive,
$q$ negative and $n-p-q$ zero eigenvalues) we define its {\it index} 
${\rm ind}(G):=p-q$. Then, combining (\ref{a.3}) and (\ref{a.5}) we 
obtain
\be \label{a.6}
I=\frac{\sqrt{2\pi}^n\; e^{-i\frac{i}{2} j^T G^{-1} 
j}\;e^{i{\rm ind}(G)\pi/4}}{|\det(G)|}
\ee

\section{On the index of special matrices}
\label{sb}

While determinants maybe tedious to calculate, its is always 
analytically possible. However, the index of a matrix is harder to 
obtain. While there exist algorithms to obtain it just from its 
characteristic polynomial (rather than from its spectrum which would be 
impossible to determine analytically for general large matrices) for 
{\it 
concrete matrices} such as D\'escartes sign rule \cite{27}, for general 
matrices of a given restricted structure
there are no such algorithms available except for in a few cases.

Our situation is the following: Consider the 16 x 16 matrix $G$ 
with entries $G_{(\mu I),(\nu J)}:=G^{\mu\nu}_{IJ}$. Since 
$G^{\mu\nu}_{IJ}=-G^{\nu\mu}_{IJ}= 
-G^{\mu\nu}_{JI}=G^{\nu\mu}_{JI}$ it is symmetric. Let us also write  
$e^{\mu I}:=e^I_\mu$. We consider the lexicographic ordering of the 
compound index $(\mu I)$ as 
$(00),..,(03),(10),..,(13),(20),..,(23),(30),..,(33)$. Consider the 
antisymmetric 4 x 4 matrix $G^{\mu\nu}$ with $0\le \mu<\nu\le 3$ given
by $(G^{\mu\nu})_{IJ}:=G^{\mu\nu}_{IJ}$. Then the 16 x 16 matrix $G$
has the following block structure
\be \label{b.1}
G = \left( \begin{array}{cccc}
0 & G^{01} & G^{02} & G^{03} \\
-G^{01} & 0 & G^{12} & G^{13} \\
-G^{02} & -G^{12} & 0 & G^{23} \\
-G^{03} & -G^{13} & -G^{23} & 0 
\end{array}
\right)
\ee
This leads us to the following conjecture.
\begin{Conjecture} \label{con1} ~\\
Let $A,B,C,D,E,F$ be antisymmetric, real valued 4 x 4 matrices 
and let $G$ be the symmetric 16 x 16 matrix
\be \label{b.2}
G = \left( \begin{array}{cccc}
 0 &  A &  B  & C \\
-A &  0 &  D  & E \\
-B & -D &  0  & F \\
-C & -E & -F  & 0 
\end{array}
\right)
\ee
Then ind$(G)=0$.
\end{Conjecture}
It turns out to be extremely hard to prove this conjecture although it 
is rather plausible. For example, it is easy to show that the conjecture 
is correct when the matrices $A,B,C,D,E,F$ are 2 x 2 antisymmetric 
matrices. It is also true when the matrices $A,B,C,D,E,F$ are linearly 
dependent. We delay the proof (or disproof) of this conjecture to future 
publications. 

If the conjecture was true and $G$ is non singular then we would know 
that $G$ has eight positive and eight negative eigenvalues. Hence 
we would know that $\det(G)>0$. In order to compute $\det(G)$ we make 
use 
of the following basic factorisation property for an arbitrary block 
matrix with blocks $A,B,C,D$ with invertible $A$
\be \label{b.3}
\left( 
\begin{array}{cc}
A & B \\
C & D
\end{array}
\right)
=
\left( 
\begin{array}{cc}
A & 0 \\
C & 1
\end{array}
\right)
\;
\left( 
\begin{array}{cc}
1 & A^{-1} B \\
0 & D-C A^{-1} B
\end{array}
\right)
\ee
It follows that
\be \label{b.4}
\det(G)=\det(A)\;\det(D-C A^{-1} B)
\ee
In our situation, by means of (\ref{b.4}) we can iteratively downsize 
the size of the matrix of which
we have to compute the determinant from rank 16 to 8 and then to 4. 
At rank 4 we may use Cayley's theorem \cite{27} in order to express 
$\det(G)$ directly in tems of polynomials of the traces of products 
of the $G^{\mu\nu}$
and thus in terms of traces of products of the plaquette loops.

\end{appendix}

\end{document}